\pdfoutput=1
\RequirePackage{ifpdf}
\ifpdf 
\documentclass[pdftex]{sigma}
\else
\documentclass{sigma}
\fi

\usepackage{mathtools}

\usepackage{bbm}
\usepackage{braket}
\usepackage{stmaryrd}
\usepackage{makecell}

\numberwithin{equation}{section}

\newtheorem{Theorem}{Theorem}[section]
\newtheorem*{Theorem*}{Theorem}
\newtheorem{Corollary}[Theorem]{Corollary}
\newtheorem{Lemma}[Theorem]{Lemma}
\newtheorem{Proposition}[Theorem]{Proposition}

\theoremstyle{definition}
\newtheorem{Definition}[Theorem]{Definition}

\newtheorem{Remark}[Theorem]{Remark}

\begin{document}


\newcommand{\arXivNumber}{2508.20797}

\renewcommand{\PaperNumber}{015}

\FirstPageHeading

\ShortArticleName{Higher-Order Linear Differential Equations for Unitary Matrix Integrals}

\ArticleName{Higher-Order Linear Differential Equations \\ for Unitary Matrix Integrals: \\ Applications
 and Generalisations \\ {\normalsize (with an Appendix by Folkmar Bornemann)}}

\Author{Peter J. FORRESTER~$^{\rm a}$ and Fei WEI~$^{\rm b}$}

\AuthorNameForHeading{P.J.~Forrester and F.~Wei}

\Address{$^{\rm a)}$~School of Mathematics and Statistics, The University of Melbourne,
Victoria 3010, Australia}
\EmailD{\mail{pjforr@unimelb.edu.au}}

\Address{$^{\rm b)}$~Department of Mathematics, University of Sussex, Brighton, BN1 9RH, UK}
\EmailD{\mail{weif0831@gmail.com}}

\ArticleDates{Received September 04, 2025, in final form January 26, 2026; Published online February 18, 2026}

\Abstract{In this paper, we consider characterisations of the class of unitary matrix integrals \smash{$\big\langle (\det U)^q {\rm e}^{s^{1/2} \operatorname{Tr}(U + U^\dagger)} \big\rangle_{U(l)}$} in terms of a first-order matrix linear differential equation for a~vector function of size $l+1$, and in terms of a scalar linear differential equation of degree~${l+1}$. It will be shown that the latter follows from the former. The matrix linear differential equation provides an efficient way to compute the power series expansion of the matrix integrals, which with $q=0$ and $q=l$ are of relevance to the enumeration of longest increasing subsequences for random permutations, and to the question of the moments of the first and second derivative of the Riemann zeta function on the critical line, respectively. This procedure is compared against that following from known characterisations involving the $\sigma$-Painlev\'e III$'$ second-order nonlinear differential equation. We show too that the natural~$\beta$ generalisation of the unitary group integral permits characterisation by the same classes of linear differential equations.}


\Classification{15B52; 37K10; 05A15}

\section{Introduction}

\subsection[A unitary matrix integral and the longest increasing subsequence problem]{A unitary matrix integral and the longest increasing\\ subsequence problem}

The study of this article is the systematic derivation of linear differential equations satisfied by
\begin{equation}\label{1.0}
\big \langle {\rm e}^{z \operatorname{Tr}(U + U^\dagger)} \big \rangle_{U(l)},
\end{equation}
the same average with a further factor of $(\det U)^q$, and their $\beta$ generalisation. In \eqref{1.0}, $U(l)$~refers to the group of $l \times l$ unitary matrices sampled with Haar measure. This matrix integral first appeared in the theoretical physics literature on two-dimensional Yang--Mills gauge theory as a~partition function \cite{Mi75}. However, it is not in this context that interest in differential equations for matrix integrals first arose. Rather, it came about due to the realisation that the coefficients~$\{T_l(N)\}$ say of $z^{2N}/(N!)^2$ in the power series of \eqref{1.0} (note that this is even in $z$) are for a given $l$ an integer family, e.g.,~for $l=4$ and starting at $N=1$
\cite[sequence A047889]{OEIS}
\begin{gather}\label{1.1}
1,2,6,24,119,694,4582,33324,261808,2190688,19318688,
178108704,\dots\,.
\end{gather}
Thus, according to \cite{Ra98}, $T_l(N)$ counts the number or permutations of $\{1,2,\dots,N\}$ such that the length of the longest increasing subsequence is less than or equal to $l$
(this number is $N!$ for $N \le l$ as illustrated in \eqref{1.1}).
{An alternative specification of $T_l(N)$, used in, e.g.,~\cite{BG00}, is
as the number of pairs of standard tableaux of content $N$ with length of the first row less than or equal to $l$}. For an explanation of these notions, and the equivalence of the two descriptions, see, e.g.,~\cite{Fu97,Ro15}.

In a different guise, a formula equivalent to
\begin{equation}\label{1.2}
\big \langle {\rm e}^{z \operatorname{Tr} (U + U^\dagger)} \big \rangle_{U(l)} =
1 + \sum_{N=1}^\infty \frac{T_l(N)}{(N!)^2} z^{2N}
\end{equation}
appeared in the earlier work \cite{Ge90}. There it was shown
\begin{equation}\label{1.3}
\det [ I_{j-k}(2z) ]_{j,k=1,\dots,l} =
1 + \sum_{N=1}^\infty \frac{T_l(N)}{(N!)^2} z^{2N},
\end{equation}
where $I_\nu(x) = {1 \over 2 \pi} \int_0^{2 \pi} {\rm e}^{x \cos \theta} {\rm e}^{{\rm i} \nu \theta} {\rm d }\theta$, $\nu \in \mathbb Z$, is the modified Bessel function of the first kind (i.e., the I-Bessel function). The reason why the Toeplitz determinant in \eqref{1.3} is equivalent to the matrix integral in \eqref{1.2} follows from the general identities
\begin{align}
\left \langle \prod_{j=1}^l a(\theta_j) \right \rangle_{U(l)} &=
{1 \over (2 \pi)^l l!} \int_0^{2 \pi} {\rm d} \theta_1 \cdots \int_0^{2 \pi} {\rm d} \theta_l \prod_{j=1}^l a(\theta_j)
\prod_{1 \le j < k \le l} \big| {\rm e}^{{\rm i} \theta_k} - {\rm e}^{{\rm i} \theta_j} \big|^2\nonumber \\
&= \det \left [ {1 \over 2 \pi} \int_0^{2 \pi} a(\theta) {\rm e}^{{\rm i} (j-k) \theta}
{\rm d }\theta \right ]_{j,k=1,\dots,l}.\label{1.4}
\end{align}
Here the first equality follows from Weyl's result on the explicit functional form of the eigenvalue probability density function (PDF) for matrices from $U(l)$ with Haar measure \cite{DF17,We39}, while the second may be regarded as a circular analogue of Andr\'eief's identity for the integral over the product of two determinants \cite{Fo19}, with use too of the Vandermonde determinant formula to rewrite the product over the differences squared in this form.

Arguments were given in \cite{Ge90} that the generating function in \eqref{1.2} is $D$-finite, meaning that it must satisfy a linear differential equation with polynomial coefficients;
for more on this see Appendix~\ref{appendixB1}. Later, in \cite{BG00}, these differential equations where determined explicitly for $l \le 7$. For example, in the case $l=4$ and thus relating to \eqref{1.1}, one has that \eqref{1.2} with $z \mapsto \sqrt{z}$ is a~solution of the fifth-order equation \cite[last collection of displayed equations]{BG00}
\begin{gather}
z^4 Y^{(5)} + 20 z^3 Y^{(4)} - 2 (10z - 59) z^2 Y^{(3)} - 2 (91 z -110) z Y'' \nonumber
\\
\qquad + 4 (16z^2 -87 z + 20) Y' + 16 (8z - 5)Y = 0\label{1.5}
\end{gather}
(there is also a known linear differential equation of degree four for the power series formed according to the rule that the coefficient of $z^k$ is given by the $k$-th member of \eqref{1.1}, and the coefficient of $z^0$ is unity \cite[Section~3.3]{B+20}).
In the present paper, using the theory of Selberg correlation integrals (see \cite{Ao87, Da68, FR12} and \cite[Section~4.6]{Fo10}), we will show how to systematically obtain a first-order $(l+1) \times (l+1)$ matrix linear differential equation, which has as the first component of its vector solution equal to
\eqref{1.2}.
As a preliminary step, the matrix average is expressed in terms of a certain Jack polynomial based hypergeometric function of $l$ variables, specialised to the case that all the variables are equal; see Section \ref{sec2}. For a closely related class of hypergeometric functions, earlier work \cite{FK22} has identified a matrix differential equation characterisation. It is shown in Section \ref{sec3} how the Jack polynomial based hypergeometric functions of present interest can be deduced as confluent limits of those appearing in \cite{FK22}, allowing for these too to be
similarly characterised.
This implies a scalar linear differential equation of degree $l+1$ satisfied by \eqref{1.2} (Proposition~\ref{P3.3} below), which with the use of computer algebra can be made explicit. As an illustration, we do this explicitly for $l=8$, which is the first case beyond the listing in \cite{BG00}; see Proposition~\ref{P3.4} below.

Efficient methods for the computation of $\{T_l(N)\}$ have come to the fore in the recent works~\cite{Bo24,FM23} (see too \cite{Bo24a}), where a primary aim was to study a scaling regime in the neighbourhood of $l=2 \sqrt{N}$ for $N$ large. A precise limit law for the corresponding scaled distribution has been known since \cite{BDJ99}, and these recent works set about quantifying correction terms. With the numerical tabulations achieved in these works, clear functional forms of corrections were able to exhibited graphically.
Such computations were carried out using the known relation for
\eqref{1.0} and a particular
$\sigma$-Painlev\'e
III$'$ equation \cite{FW02,TW94b}, and also its Chazy-I equivalent form.
In Appendix \ref{appB2}, a comparative discussion of the computational complexity associated with the latter characterisation,
and that based on the present matrix differential equation characterisation via the vector recurrence \eqref{MD3a0}, is given.

\subsection[A unitary matrix integral and the moments of the second derivative of the Riemann zeta function on the critical line]{A unitary matrix integral and the moments of the second derivative\\ of the Riemann zeta function on the critical line}

{We have a second main motivation for considering the characterisation of the matrix integral~\eqref{1.0}, and that of modifications of this integral, in terms of
a linear differential equation equations satisfied by~\eqref{1.0}}.
This is the primary role played by the modification of \eqref{1.0} and the Toeplitz determinant in \eqref{1.3}
\begin{align}
\big \langle (\det U)^{l} {\rm e}^{z \operatorname{Tr}(U + U^\dagger)} \big \rangle_{U(l)}& = \det [ I_{j-k+l}(2z) ]_{j,k=0,\dots,l-1}\nonumber\\
&=
(-1)^{l(l-1)/2} \det [ I_{j+k+1}(2z) ]_{j,k=0,\dots,l-1} \label{1.6}
\end{align}
in the recent studies \cite{KW24a, KW24b}.
Here the second equality views the determinant in the Hankel class (i.e.,~indices a function of $j+k$), which is the structure emphasised in these studies.
The theme of \cite{KW24a, KW24b} is the question of statistical properties, particularly moments, of higher-order derivatives of the Riemann zeta function on the critical line.
Fundamentally, in keeping with the celebrated work \cite{KS00}, it is hypothesised that these zeta function moments coincide with the scaled limiting moments of the corresponding derivative of the characteristic polynomial of a~Haar distributed random unitary matrix;
see \cite{ABGS21,AKW22,B+19a,B+19, CRS06,De08,Fo22a,Hu01,Wi12} for earlier work in this direction, and
\cite{A+24,AGKW24,A+25,BW25,SW24} for subsequent developments.
In particular, in relation to the second derivative, this leads to the study of
\cite{KW24a}
\begin{equation}\label{1.8}
F_{2}(l)=\lim_{N\rightarrow \infty}\frac{1}{N^{l^2+4l}}\big \langle |Z_{N}''(1)|^{2l} \big \rangle_{U(N)},
\end{equation}
where
\[
Z_{N}(s) := {\rm e}^{-\pi {\rm i }N/2}{\rm e}^{{\rm i} \sum_{n=1}^{N}\theta_{n}/2}
s^{- N/2}\prod_{j=1}^{N}\bigl(1-s{\rm e}^{-{\rm i}\theta_{j}}\bigr).
\]
In \cite{KW24a}, it was shown
\[
F_{2}(l)=(-1)^{\frac{l(l-1)}{2}}
\sum_{m=0}^{2l} \binom{2l}{m}
\left( \frac{\rm d}{{\rm d}x} \right)^{4l-2m} \bigl( {\rm e}^{-\frac{x}{2}} x^{-m-\frac{l^2}{2} } f_{m}(x) \bigr) \big|_{x= 0},
\]
where each \( f_m(x) \) is a combinatorial linear combination of shifted versions of the Hankel determinant appearing on the left-hand side of \eqref{1.6}, which takes the form
\[
\det\big[I_{j+k+1+2\mu_{j+1}}\bigl(2\sqrt{x}\bigr)\big]_{j,k=0,\ldots,l-1},
\]
and where \( \mu_1 + \cdots + \mu_l = m \). Moreover, in \cite{KW24b}, it was shown to be given in the form
\[
f_{m}(x)=\frac{1}{x^{m}}\sum_{j=0}^{m}x^{j}P_{j}(x)\frac{{\rm d}^{j}\tau_{l}(x)}{{\rm d}x^{j}}.
\]
Here, each \( P_j(x) \) is a polynomial defined recursively, and \( \tau_l(x) \) denotes the Hankel determinant appearing on the left-hand side of \eqref{1.6}, with \( z = 2\sqrt{x} \).

In \cite{KW24b}, certain reasoning is given that leads to the conclusion that like \eqref{1.0}, $\tau_{l}(x)$ satisfies a linear differential equation of degree $l+1$ (see Appendix~\ref{appendixA} for an overview).
As mentioned in the opening paragraph,
our approach to systematically deriving a linear differential equation that \eqref{1.0} satisfies allows for an extra factor $(\det U)^q$ to be included for general $q$, and thus
includes \eqref{1.6}. Thus in this regards our work complements \cite{KW24b}. We consider this further in Section~\ref{S4.3}, and discuss too the advantage of the linear differential equation characterisation over an alternative nonlinear
differential equation characterisation put forward in \cite{FW06}.

\section{Jack polynomial based hypergeometric function form}\label{sec2}

It has been revised in the introduction that the matrix integral \eqref{1.0} has the interpretations as a partition function for a two-dimensional gauge theory, and as the generating function for counting longest increasing subsequences. There is also an occurrence of \eqref{1.0} as a limiting gap probability at the hard edge ({neighbourhood of the origin}) of a positive definite matrix ensemble with unitary symmetry. To specify this, define complex Wishart matrices $W = X^\dagger X$ with $X$ an $n \times N$ $(n \ge N)$ standard complex Gaussian matrix (also referred to as a rectangular GinUE matrix \cite{BF25}). Let {$E_N((0,s);l)$} with $l = n - N$ denote the probability of no eigenvalues in $(0,s)$
{in this class of complex Wishart ensembles.}
In \cite{Fo93c}, it was established that
\begin{equation}\label{2.0}
E_{\beta = 2}^{\rm hard}((0,s);l) :=
\lim_{N \to \infty} E_N((0,s/(4N));l) =
{\rm e}^{-s/4}
\big \langle {\rm e}^{(s^{1/2}/2) \operatorname{Tr}(U + U^\dagger)} \big \rangle_{U(l)}.
\end{equation}
{(Here the use of the symbol $\beta$ as a subscript on the left-hand side is in the context of the underlying Dyson index; see \eqref{CB} below and subsequent text.)}
The most significant contribution in \cite{Fo93c} for present purposes was the identification of the right-hand side of \eqref{2.0}
(and generalisations) as a Jack polynomial based generalised hypergeometric function.

To specify this class of multivariable special function, we first recall that, with $\mathbf x = (x_1,\dots,\allowbreak x_n)$
the Jack polynomials \smash{$P_\kappa^{(\alpha)}(\mathbf x)$} are symmetric polynomials of $n$ variables \cite[Chapter~VI.6]{Ma95}, \cite[Chapter~12]{Fo10}, \cite[Chapter~7]{KK09}.
 They are dependent on
 a parameter $\alpha > 0$ and
 labelled by a~partition $\kappa :=
 (\kappa_1,\dots,\kappa_n)$; in the latter
 each part $\kappa_i \in \mathbb Z_{\ge 0}$ is ordered as $\kappa_1 \ge \kappa_2 \ge \cdots \ge \kappa_n$. These polynomials are eigenfunctions of
 the
differential operator
\begin{equation}\label{4.0}
\sum_{j=1}^n
x_j^2 {\partial^2 \over \partial x_j^2}
+ {2 \over \alpha} \sum_{1 \le j < k \le n}{1 \over x_j -
x_k}
\left (x_j^2 {\partial \over \partial x_j} - x_k^2 {\partial
\over
\partial x_k}
\right ).
\end{equation}
 Introducing the dominance partial ordering on partitions of the same length
 $\mu < \kappa$, the Jack polynomial
\smash{$P_\kappa^{(\alpha)}(\mathbf x)$} is uniquely determined as the eigenfunction of \eqref{4.0} having
the triangular structure with respect to the monomial basis $\{ m_\mu(\mathbf x) \}$,
\smash{$
P_\kappa^{(\alpha)}(\mathbf x) = m_\kappa(\mathbf x) +
\sum_{\mu < \kappa} c_{\kappa, \mu}^{(\alpha)}
m_\mu (\mathbf x)$}.

In the definition of Jack polynomial based hypergeometric functions, the Jack polynomials play the role of the monomials, although these must be scaled
\smash{$
P_\kappa^{(\alpha)}(\mathbf x) \mapsto \alpha^{|\kappa|} P_\kappa^{(\alpha)}(\mathbf x)/ h_\kappa'$}.
{Here} the precise definition of $h_\kappa'$
(which depends too on $\alpha$) can be found in, e.g.,~\cite[equation~(12.46)]{Fo10}.
Playing the role of the Pochhammer symbol is the quantity
 \[
\qquad[u]_\kappa^{(\alpha)} := \prod_{j=1}^n {\Gamma(u - (j-1)/\alpha + \kappa_j) \over \Gamma(u - (j - 1)/\alpha)};
\]
see, e.g.,~\cite[equation~(12.46)]{Fo10}. With this notation we then have as the series definition of the generalised hypergeometric function in question (see, e.g.,~\cite[Section~13.1]{Fo10})
\[
 {\vphantom{F}}_p^{\mathstrut} F_q^{(\alpha)}(a_1,\dots,a_p;b_1,\dots,b_q; \mathbf x):=\sum_\kappa {\alpha^{| \kappa |} \over h_\kappa'}
 \frac{[a_1]^{(\alpha)}_\kappa\dots [a_p]^{(\alpha)}_\kappa }{[b_1]^{(\alpha)}_\kappa
\dots [b_q]^{(\alpha)}_\kappa}
P_\kappa^{(\alpha)}(\mathbf x).
\]
In the case $n=1$, this is independent of $\alpha$ and reduces to its classical scalar counterpart ${}_pF_q$.

We are now in a position to specify the result of \cite{Fo93c} which allows
the right-hand side of~\eqref{2.0} to be identified
as a Jack polynomial based generalised hypergeometric function. Thus, for general~${\alpha > 0}$,
\begin{align}
{\vphantom{F}}_0^{\mathstrut} F_1^{(\alpha)}(c + (n - 1)/\alpha;(t)^n)
={}& B_n(c,\alpha) \left ( {1 \over t} \right )^{(c-1)n/2}\left ( {1 \over 2 \pi } \right )^n \int_{[-\pi,\pi]^n} \prod_{j=1}^n
{\rm e}^{2t^{1/2} \cos \theta_j} {\rm e}^{{\rm i} (c-1) \theta_j}
\nonumber \\
&
\times \prod_{1 \le j < k \le n}
\big|{\rm e}^{{\rm i} \theta_k} - {\rm e}^{{\rm i} \theta_j} \big|^{2/\alpha} {\rm d}\theta_1 \cdots {\rm d}\theta_n,\label{15.141x}
\end{align}
(see too \cite[equation~(13.27)]{Fo10} and \cite[equation~(4.32)]{Fo25}),
where $c \in \mathbb Z_{\ge 1}$ (this restriction can be removed by suitably deforming the implied circular contours in \eqref{15.141x} to Hankel loops \cite[proof of Proposition~2]{Fo13}) and
\[
B_n(c,\alpha) = \prod_{j=1}^n {\Gamma(1 + 1/\alpha) \Gamma(c + (j-1)/\alpha)
\over \Gamma(1 + j/\alpha) }.
\]
The notation $(t)^n$ on the left-hand side denotes that $\mathbf x = (t,t,\dots,t)$, and thus all $n$ components are equal to $t$.
Making use of \eqref{1.4}, it follows from \eqref{15.141x} that
\begin{equation}\label{2.8}
\big \langle {\rm e}^{s^{1/2} \operatorname{Tr}(U + U^\dagger)} \big \rangle_{U(l)} =
{\vphantom{F}}_0^{\mathstrut} F_1^{(\alpha)}\bigl(l;(s)^l\bigr) \big |_{\alpha = 1};
\end{equation}
for a derivation of this result, and its generalisation for $\alpha=1/2$ and $\alpha = 2$, based on the theory of zonal polynomials, see \cite{FR09}. Extending this, for any $q \in \mathbb Z_{\ge 0}$ we read off that
\begin{equation}\label{2.8a}
\big \langle (\det U)^q {\rm e}^{s^{1/2} \operatorname{Tr}(U + U^\dagger)} \big \rangle_{U(l)} =
{s^{q l/2} \over l! B_l(q+1,\alpha)}
{\vphantom{F}}_0^{\mathstrut} F_1^{(\alpha)}\bigl(q+l;(s)^l\bigr) \big |_{\alpha = 1};
\end{equation}
recall \eqref{1.6} in relation to the case $q=l$.

Specifying the circular $\beta$ ensemble CE${}_{\beta,n}$ is the eigenvalue PDF
\begin{equation}\label{CB}
{1 \over (2 \pi)^n C_{\beta,n}}
\prod_{1 \le j < k \le n}
\big|{\rm e}^{{\rm i} \theta_k} - {\rm e}^{{\rm i} \theta_j} \big|^\beta, \qquad C_{\beta,n} = {\Gamma(\beta n/2+1) \over \Gamma(\beta/2+1)^n};
\end{equation}
the case $\beta =2$ corresponds to $U(n)$.
This gives the $\beta$ generalisation of \eqref{2.8a}
\begin{equation}\label{2.9a}
\big \langle (\det U)^q {\rm e}^{s^{1/2} \operatorname{Tr}(U + U^\dagger)} \big \rangle_{{\rm CE}_{\beta,l}} =
{s^{q l/2} \over C_{\beta,l} B_l(q+1,2/\beta)}
{\vphantom{F}}_0^{\mathstrut} F_1^{(2/\beta)}\bigl(q+1 + (l-1)\beta/2 ;(s)^l\bigr).
\end{equation}
For the $\beta$-generalisation of the eigenvalue PDF for complex Wishart matrices, one has the functional form proportional to
\[
\prod_{l=1}^N x_l^a {\rm e}^{-\beta x_l/2}
\mathbbm 1_{x_l > o} \prod_{1 \le j < k \le N} |x_k - x_j|^\beta;
\]
see, for example, \cite[equation~(3.16) with $a$ appropriately identified]{Fo10}. For this ensemble, let $E_{N,\beta}((0,s); a)$ denote the probability of there being no eigenvalues in $(0,s)$. Generalising the equality implied by \eqref{2.0} and \eqref{2.8}, it was shown in \cite{Fo93c} that for $a \in \mathbb Z_{\ge 0}$
\[
E_{\beta}^{\rm hard}((0,s);a) :=
\lim_{N \to \infty} E_{N,\beta}((0,s/(4N));a) =
{\rm e}^{-\beta s}
{\vphantom{F}}_0^{\mathstrut} F_1^{(\beta/2)}(2 a/\beta ;(s/4)^a).
\]
This then relates to \eqref{2.9a} with $\beta \mapsto 4/\beta$,
$l=a$, $q=2/\beta - 1$.

\section{Matrix differential equation characterisation}\label{sec3}

As in the classical case, there is the limit relation
\begin{equation}\label{2.8b}
\lim_{M \to \infty}
{\vphantom{F}}_1^{\mathstrut} F_1^{(\alpha)}(a M;b;\mathbf x/M) =
{\vphantom{F}}_0^{\mathstrut} F_1^{(\alpha)}(b;\mathbf x).
\end{equation}
The significance of this is that for $\mathbf x = (t)^n$, since the 2022 work
\cite{FK22} we have available a matrix differential equation characterisation of
\smash{${\vphantom{F}}_1^{\mathstrut}F_1^{(\alpha)}(a ;b;\mathbf x)$}.
This comes about in the setting of the integral representation \cite[Corollary~2.1]{Fo10}
\begin{gather}
 {\vphantom{F}}_1^{\mathstrut} F_1^{(2/\beta)} ( (\beta/2) (n-1) + a + 1 ; \beta (n - 1) + a + b + 2 ;(-s)^n ).\nonumber\\
\qquad={1 \over S_n(a,b,\beta)} \int_0^1 {\rm d}x_1 \cdots \int_0^1 {\rm d}x_n
\prod_{l=1}^n x_l^{a} (1 - x_l)^{b} {\rm e}^{-s x_l} \prod_{1 \le j < k \le n} | x_k - x_j |^\beta,\label{3.2}
\end{gather}
valid for Re$(a) > -1$, Re$(b) > -1$, {Re$(\beta) \ge 0$}. Here $S_n(a,b,\beta)$ is the Selberg integral (see, e.g.,~\cite[Chapter~4]{Fo10}), which normalises the right-hand side so that it is equal to unity for $s=0$, and we further remark that the integral is well defined too in a range of negative $\beta$ values, although this detail is not used in our work. The matrix differential equation, which we will first present in the form of a differential-difference recurrence, relates to the family of multi-dimensional integrals
\begin{gather*}
H_{p}(x) = {1 \over C_p^n}\int_0^1 {\rm d}t_1 \cdots \int_0^1 {\rm d}t_n
\prod_{l=1}^n t_l^{a} (1 - t_l)^{b-1} {\rm e}^{-x \sum_{l=1}^n t_l} \\
\phantom{H_{p}(x) = }{}\times \prod_{1 \le j < k \le n} | t_k - t_j|^\beta
e_p(1 - t_1,\dots, 1 - t_n),
\end{gather*}
for $p=0,1,\dots,n$.
The factor $e_p(1 - t_1,\dots, 1 - t_n)$ in the integrand denotes the $p$-th elementary symmetric polynomials in the variables $\{ (1 - t_i) \}_{i=1}^n$, while
$C_p^n$ is the binomial coefficient $n$ choose $p$. Setting $p=n$ gives the right-hand side of
\eqref{3.2} and thus \smash{${}^{}_1 F^{(2/\beta)}_1$}, up to the Selberg integral normalisation, and setting $p=0$ gives the same with $b$ replaced by $b-1$.

\begin{Proposition}[{\cite[Corollary~2.5]{FK22}}] \label{P3.1}
The multi-dimensional integrals $\{ H_p(x) \}_{p=0}^{n}$ satisfy the
differential-difference system
\begin{gather}
(n - p) x H_{p+1}(x)
= ( (n - p) x + B_p ) H_p(x) + x {{\rm d} \over {\rm d}x} H_p(x) - D_p H_{p-1}(x), \nonumber\\
 p=0,\dots, n,\label{6.1a}
\end{gather}
where
\[
B_p = p ( a + b + 1 + (\beta/2)(2n - p - 1) ), \qquad
D_p = p ( (\beta/2) (n - p) + b ).
\]
\end{Proposition}

\begin{Corollary}
 Introduce the column vector $\mathbf H(x) = [H_p(x)]_{p=0}^n$, and the bi-diagonal matrices
\begin{align*}
\mathbf X & = - \operatorname{diag} [n, n - 1,\dots,0] + \operatorname{diag}^+ [n,n-1,\dots,1] , \\
\mathbf Y & = - \operatorname{diag} [B_0, B_1,\dots, B_n] + \operatorname{diag}^- [D_1, D_2,\dots, D_n].
\end{align*}
Here $\operatorname{diag}^+$ {\rm (}$\operatorname{diag}^-${\rm )} denotes the leading diagonal above $($below$)$ the diagonal itself.
We have that the differential-difference equation \eqref{6.1a} is equivalent to the matrix differential equation
\begin{equation}\label{MD}
x {{\rm d} \over {\rm d}x} \mathbf H(x) = (x \mathbf X + \mathbf Y) \mathbf H(x).
\end{equation}
As a consequence {\rm\cite[\emph{Remark} 4.19]{FK22}}, introducing the vector of power series solutions
\[
 \mathbf H(x) = \left [ \sum_{k=0}^\infty c_{p,k} x^k \right ]_{p=0}^n,
\]
 we have
\begin{gather*}
 c_{0,0} = S_n(a,b-1,\beta), \qquad c_{n,0} = S_n(a,b,\beta), \qquad
 c_{n - p,0} = c_{n,0} \prod_{s=1}^p {B_{n+1-s} \over D_{n+1-s}}, \\ p=1,\dots,n
\end{gather*}
 and, with $\mathbf c_k := [ c_{p,k} ]_{p=0}^n$, the vector recurrence relation
 \begin{equation}\label{MD3}
 (k {\mathbf I}_{n+1} - \mathbf Y) \mathbf c_k = \mathbf X \mathbf c_{k-1}, \qquad\mathbf c_{-1} := \mathbf 0,\qquad k=1,2,\dots,
 \end{equation}
 where $\mathbf{I}_{n+1}$ denotes the $(n+1) \times (n+1)$ identity.
\end{Corollary}

We now want to make use of \eqref{2.8b} in
Proposition~\ref{P3.1}, in view of both the first and last components of $\mathbf H(x)$ relating to \smash{${}^{}_1 F^{(2/\beta)}_1$}.

\begin{Proposition}\label{P3.2}
 Consider the
differential-difference system\footnote{Here and in the remainder of the paper the functions $\{I_p(x) \}$ are not to be confused with
the I-Bessel function as appearing in~\eqref{1.3} and~\eqref{1.6}.}
\begin{equation}\label{6.1b}
 (n - p) x I_{p+1}(x)
= B_p I_p(x) + x {{\rm d} \over {\rm d}x} I_p(x) - p I_{p-1}(x), \qquad p=0,\dots, n,
\end{equation}
where
$
B_p$ is as in \eqref{6.1a}. Introduce the column vector
${\mathbf I}(x) = [I_p(x)]_{p=0}^n$, and the matrices
\[
{\mathbf W} = \operatorname{diag}^+ [n,n-1,\dots,1], \qquad
{\mathbf Z} = - \operatorname{diag} [B_0, B_1,\dots, B_n] + \operatorname{diag}^- [1, 2,\dots, n].
\]
We have that the differential-difference equation \eqref{6.1b} is equivalent to the matrix differential equation
\begin{equation}\label{MDa1}
x {{\rm d} \over {\rm d}x}\mathbf I(x) = (x\mathbf W +\mathbf Z) \mathbf I(x).
\end{equation}
Furthermore, introducing the vector of power series solutions
\begin{equation}\label{3.10a}
\mathbf I(x) = \left [ \sum_{k=0}^\infty d_{p,k} x^k \right ]_{p=0}^n,
\end{equation}
normalised by choosing $d_{0,0} =1$,
we have
\begin{align}
I_0(x) & = {\vphantom{F}}_0^{\mathstrut} F_1^{(2/\beta)} ( \beta (n - 1) + a + b + 1 ;(x)^n ),
\label{MDa22} \\
I_n(x) & = d_{n,0} {\vphantom{F}}_0^{\mathstrut} F_1^{(2/\beta)} ( \beta (n - 1) + a + b + 2 ;(x)^n )\label{MDa2},
\end{align}
and, provided all the $B_s$ are nonzero,
 \begin{equation}\label{MDa2x}
 d_{p,0} = {p! \over \prod_{s=1}^p B_s},
 \qquad p=1,\dots,n.
\end{equation}
 The coefficient vector $\mathbf d_k := [ d_{p,k} ]_{p=0}^n$ satisfies the recurrence relation
 \begin{equation}\label{MD3a0}
 (k {\mathbf I}_{n+1} -\mathbf Z) \mathbf d_k =\mathbf W \mathbf d_{k-1},\qquad k=1,2,\dots.
 \end{equation}
\end{Proposition}

\begin{proof}
To deduce \eqref{6.1b} and \eqref{MDa1}, we substitute in \eqref{6.1a} and \eqref{MD}, respectively, $x \mapsto -x/a$, $H_p(-x/a) = b^p I_p(x)$ and take the limit $a \to \infty$ with $a+b$ fixed.
The evaluations \eqref{MDa2} follow from the evaluations of $H_n(x)$,
$H_0(x)$ in terms of \smash{$\vphantom{F}_1^{\mathstrut} F_1^{(2/\beta)}$} noted in the sentence above Proposition~\ref{P3.1}.
The evaluation \eqref{MDa2x} can be calculated as the eigenvector of ${\mathbf Z}$ corresponding to the eigenvalue 0, normalised to have its first component equal to unity. The vector recurrence relation \eqref{MD3a0} follows by substituting \eqref{3.10a} in \eqref{MDa1} and equating like powers of $x$.
\end{proof}

Analogous to the situation with \eqref{MD3},
since ${\mathbf d}_0$ is known from \eqref{MDa2x} the vector recurrence~\eqref{MD3a1} uniquely determines
 ${\mathbf d}_1,{\mathbf d}_2,\dots$ provided $(l {\mathbf I}_{n+1} -\mathbf Z)$ is invertible for all $l \in \mathbb Z^+$. Since~$\mathbf Z$ is lower triangular, the criterium for the latter is simply $- B_p \notin \mathbb Z^+$,
for each $p=1,\dots,n$. Relevant to \eqref{2.9a} are the choices in Proposition~\ref{P3.2} of
$n=l$, and $a+b$ such that
\begin{equation}\label{MD3a3}
B_p = p ( q + 1+ (\beta/2) (l - p ) ),
\end{equation}
as follows from \eqref{MDa2}.
This for $q \in \mathbb Z_{\ge 0}$ and $p=0,\dots,l$ is non-negative, allowing then for the applicability of \eqref{MD3a0} written in the form
\begin{equation}\label{MD3a1}
 \mathbf d_k = (k {\mathbf I}_{l+1} -\mathbf Z)^{-1}\mathbf W \mathbf d_{k-1},\qquad k=1,2,\dots
 \end{equation}
and with the initial condition $\mathbf d_0 = [d_{p,0}]_{p=0,\dots,n}$, where $d_{0,0}=1$ and $d_{p,0}$ ($p \ge 1$) is as given by~\eqref{MDa2x}. Further specialising to $q=0$, $\beta = 2$, we can use \eqref{MD3a1} to compute $\{T_l(N) \}_{N=1,,\dots,N^*}$ (here $N^*$ is some given largest value of $N$) according to the matrix product formula
\begin{equation}\label{MD3a2a}
T_l(N) = (N!)^2
( \mathbf R_{N,l} \mathbf d_0)_{1},
\end{equation}
where the notation $( \cdot )_{1}$ denotes the first entry in the corresponding column vector and
\[
\mathbf R_{N,l} :=
(N {\mathbf I}_{l+1} -\mathbf Z)^{-1}
{\mathbf W}
((N-1) {\mathbf I}_{l+1} -\mathbf Z)^{-1}
{\mathbf W}
\cdots
( {\mathbf I}_{l+1} -\mathbf Z)^{-1}
{\mathbf W}.
\]
 As a concrete example, implementation of \eqref{MD3a2a} with $l=4$,
and thus involving products of~${5 \times 5}$ matrices, allows for generation of the sequence \eqref{1.1}.

\begin{Remark}\label{R3.1} \:
\begin{itemize}\itemsep=0pt
\item[(1)] Suppose we replace $q$ by $q-1$ in \eqref{MD3a3}. Then setting $q=0$ we observe that
the column vector $[\delta_{p,l}]_{p=0}^{l}$
 is an eigenvector of ${\mathbf Z}$
 with $n=l$,
 allowing for the choice of initial vector corresponding to $d_{p,0} = \delta_{p,l}$ ($p=0,\dots,l$). Using this initial vector, an alternative formula to \eqref{MD3a2a} is
\[
T_l(N) = (N!)^2
( \mathbf R_{N,l} \mathbf d_0)_{l+1},\]
where the notation $( \cdot )_{l+1}$ denotes the $(l+1)$-th entry (final entry) in the corresponding column vector. In the following section, when referring to $I_n(x)$ in the case $q=0$ as it relates to \eqref{2.8}, it will be assumed that the initial vector has been similarly modified (and thus in
\eqref{MDa2} $d_{n,0}$ is replaced by unity --- without this modification
\eqref{MDa2} is invalid since~$d_{n,0}$ {as specified by \eqref{MDa2x}}
involves a division by zero).

\item[(2)] For efficient computation using the matrix differential equation, it is advantageous to forgo~\eqref{MD3a2a} in favour of a recursive computation based on \eqref{MD3a0}. See
Appendix~\ref{appB2} for a~complexity analysis when using this method to
compute $\{T_l(N)\}$ in~\eqref{1.2}, which moreover is compared against the formalism used in
\cite{Bo24,FM23} for the same purpose.
\end{itemize}
\end{Remark}

\section{Scalar differential equation}
\subsection{The case of general parameters}
Next we would like to relate \eqref{MDa1} to a scalar linear differential equation of degree $n+1$ for $I_n(x)$ (and thus by the replacement $b \mapsto b-1$
for $I_0(x)$; recall \eqref{MDa22}).

\begin{Proposition}\label{P3.3}
 Let $I_n(x)$ be specified via the differential-difference system
 \eqref{6.1b}. For some polynomials $\{ q_m(x) \}_{m=0,1,\dots,n-1}$, we have
 \begin{gather}
\left ( x^n {{\rm d}^{n+1} \over {\rm d} x^{n+1}} + q_n(x) x^{n-1} {{\rm d}^{n} \over {\rm d} x^{n}} + q_{n-1}(x) x^{n-2}{{\rm d}^{n-1} \over{\rm d} x^{n-1}}+ \cdots +
 q_2(x) x{{\rm d}^2 \over{\rm d} x^2}+
 q_1(x) {{\rm d} \over{\rm d} x} + q_0(x) \right )\nonumber\\
 \qquad\times
 I_n(x) =0.\label{MD3a2}
 \end{gather}
 The degree in $x$ of the coefficient of each
 ${{\rm d}^{j} \over{\rm d} x^{j}}$ is $[(n-1+j)/2]$, thus restricting the degrees of the~$q_m(x)$. Furthermore, when~$\{ B_j \}$
 {in \eqref{6.1b}}
 are all integers, the coefficients of these polynomials are also integers.
 \end{Proposition}

 \begin{proof}
 Let\footnote{This notation is not to be confused with $D_p$ in~\eqref{6.1a}.} $D_x := x {{\rm d} \over{\rm d} x}$.
 From \eqref{6.1b} in the case $p=n$, we read off that
 \begin{equation}\label{3.8a}
 n I_{n-1}(x) = (B_n + D_x) I_{n}(x).
 \end{equation}
 Using this in the case $p=n-1$ of \eqref{6.1b} then gives
 \begin{equation}\label{3.8b}
 n ( n - 1 ) I_{n-2}(x) = ((B_{n-1} + D_x)
 (B_{n} + D_x) - n x )
 I_{n}(x).
 \end{equation}
 Using both \eqref{3.8a} and \eqref{3.8b} in the case $p=n-2$ of
\eqref{6.1b} then gives
 \begin{gather}
 n ( n - 1 )(n-2) I_{n-3}(x)\nonumber\\
 \qquad = ( (B_{n-2} + D_x) ((B_{n-1} + D_x)
 (B_{n} + D_x) - n x )
 -2 x (n-1) (B_{n} + D_x) )
 I_{n}(x).\label{3.8c}
 \end{gather}
 We see that the differential operators on the right-hand sides of \eqref{3.8a}, \eqref{3.8b}, \eqref{3.8c} have the structure
 \begin{gather*}
 x {{\rm d} \over {\rm d}x} + c_0,\qquad
 x^2 {{\rm d}^2 \over {\rm d}x^2} + d_0 x {{\rm d} \over {\rm d}x}+
 (c_1 + c_2 x),\\
 x^3 {{\rm d}^3 \over {\rm d}x^3} + e_0 x^2 {{\rm d}^2 \over {\rm d}x^2}+
 (d_1 + d_2 x) x {{\rm d} \over {\rm d}x} +
 (c_3 + c_4x),
\end{gather*}
{for certain coefficients.}

Repeating this procedure down to the case $p=1$ of \eqref{6.1b} allows each $I_j(x)$ ($j=0,\dots,n)$ to be written as a differential operator of degree $n-j$ acting on $I_n(x)$. In particular,
$n! I_1(x)$ has the structure of a differential operator of the form \eqref{MD3a2} with $n \mapsto n - 1$ and an extra factor of~$x$ multiplying each of the terms involving a derivative. Similarly, $n! I_0(x)$ has the structure of a~differential operator of the form in
\eqref{MD3a2} as written, except for
an extra factor of $x$ multiplying each of the terms involving a derivative.

The case $p=0$ of \eqref{6.1b} gives
 \begin{equation}\label{op3a}
{{\rm d} \over{\rm d} x} I_0(x) - n I_1(x) = 0.
\end{equation}
From the established formulas for $I_0(x)$, $I_1(x)$, this implies \eqref{MD3a2}. That the coefficients of the polynomials $\{ q_j(x) \}$ are integers for $\{ B_j\}$ integers is immediate from the structure of the right-hand sides of
\eqref{3.8a}--\eqref{3.8c} as extends to $I_0(x)$, $I_1(x)$.
\end{proof}

\subsection[Characterisation of the matrix integral in (1.3)]{Characterisation of the matrix integral in (\ref{1.2})}\label{S4.2}
 In the case $\beta = 2$, $n=l$ and $B_p = p(l-p)$, $I_{n}(x)$ is according to \eqref{2.8a} and \eqref{MDa2}
 (the latter modified in keeping with Remark \ref{R3.1}.1)
 equal to the matrix average in
 \eqref{1.2} with $z = x^{1/2}$ and thus is a generating function for the sequence $\{ T_l(N) \}_{N=1,2,\dots}$.
 Using different methods, the precise form of the corresponding differential operator in \eqref{MD3a2} has been computed for $l=2,3,\dots,7$ in \cite{BG00}. From the substitution method given in the proof of Proposition~\ref{P3.3}, we can reclaim the listing of~\cite{BG00}, and extend it to higher values of $l$ (see the work \cite{RF21} for other applications in random matrix theory of this method in reducing first-order linear matrix differential equations to higher-order scalar differential equations). Here we make note of the case $l=8$.

 \begin{Proposition}\label{P3.4}
The matrix integral in \eqref{1.2} with $z = x^{1/2}$ and $l=8$ satisfies the differential equation in \eqref{MD3a2} with $n=8$ {\rm(}and thus is of degree $9)$, with
 \begin{gather*}
 q_8(x) = 120,\qquad q_7(x) = 12 (483-10 x),\qquad
 q_6(x) = 4 (36382 - 2541 x) , \\
 q_5(x) = 12 \bigl(170417 - 27414 x + 364 x^2\bigr),\qquad
 q_4(x) = 24 (214803 + 9830 x), \\
 q_3(x) = 4 \bigl(16531678 - 10189630 x + 1112127 x^2 - 13120 x^3\bigr), \\
 q_2(x) = -8 \bigl(-14762834 + 19053398 x - 4421047 x^2 + 185712 x^3\bigr), \\
 q_1(x) = 8 \bigl(5883904 - 26926706 x + 13943655 x^2 - 1483008 x^3 + 18432 x^4\bigr), \\
 q_0(x) =256 \bigl(-183872 + 380121 x - 98628 x^2 + 4608 x^3\bigr).
 \end{gather*}
 \end{Proposition}

In \eqref{1.2}, let us replace $z$ by $x^{1/2}$ so that it reads
 \begin{equation}\label{1.2e}
 \big \langle {\rm e}^{x^{1/2} \operatorname{Tr} (U + U^\dagger)} \big \rangle_{U(l)} =
1 + \sum_{N=1}^\infty {T_l(N) \over (N!)^2} x^{N}.
\end{equation}
As has just been illustrated in Proposition~\ref{P3.4}, we have from Proposition~\ref{P3.3} that \eqref{1.2e} satisfies a certain scalar linear differential equation of degree $l+1$.
As remarked in this context in \cite{BG00}, it is well known and classical that one can translate the latter into a recurrence for
$\{ T_l(N) \}_{N=1,2,\dots}$ with coefficients that are polynomial in $N$. Previous literature contains the explicit form of the difference equations for small $l$; \cite[footnote 22]{Bo24} gives references containing the explicit forms up to $l=5$. As an explicit example, from \cite{St07}, for $l=4$ we have
\begin{align}
(N+4)(N+3)^2 T_4(N) ={}& \bigl(20 N^3 + 62 N^2 +22 N - 24\bigr) T_4(N-1)\nonumber
\\& -
64 N (N-1)^2 T_4(N-2),\label{1.2X}
\end{align}
valid for $N=1,2,\dots$.
Iterating this, starting with $N=1, T_4(0) = 1$ indeed generates the sequence \eqref{1.1}. Following the proof of Proposition~\ref{P3.4e} below, note is made of the explicit forms for $l=2,3$ and 5.

From \cite{BFK95}, one has the conjecture that $\{ T_l(N) \}_{N=1,2,\dots}$ satisfy a linear difference equation of the form
\begin{equation}\label{1.2Yinaccurate} 
\sum_{k=0}^{[l/2]+1} P_k(N) T_l(N-k)=0,
\end{equation}
with each $P_k(N)$ a polynomial in $N$. However, already with $l=4$ we see from \eqref{1.2X} that \eqref{1.2Yinaccurate} is inaccurate, as \eqref{1.2X} involves only $\{ T_4(N), T_4(N-1), T_2(N-2) \}$ whereas \eqref{1.2Yinaccurate} gives too a term involving $T_2(N-3)$. Making use of
Proposition~\ref{P3.3}, we can (by inspection) give the corrected form of \eqref{1.2Yinaccurate}.

\begin{Proposition}\label{P3.4d}
 Let $\{ T_l(N) \}_{N=1,2,\dots}$ be as in \eqref{1.2e}. We have
 that this sequence satisfies a~linear difference equation of the form \eqref{1.2Y} with the upper terminal replaced by
 $[(l-1)/2] + 1$.
\end{Proposition}

Alternative to first determining the scalar differential equation and then converting it to a~difference equation, it is possible to deduce the latter directly from the differential-difference system \eqref{6.1b}.\footnote{We thank Folkmar Bornemann for this observation.}
\begin{Proposition}\label{P3.4e}
Consider the differential-difference system \eqref{6.1b}. By substituting $I_p(x) = \sum_{N=0}^\infty d_{p,N} x^N$ as is consistent with \eqref{3.10a}, a difference equation
for $\{ d_{p,N} \}_{N=0,1,\dots}$
of the form
\begin{equation}\label{1.2Y}
\sum_{k=0}^{[(n-1)/2]+1} Q_k(N) d_{n,N+1-k}=0,
\end{equation}
with each $Q_k(p)$ a polynomial in $p$, can be arrived at.
\end{Proposition}

\begin{proof}
Substituting the power series for $I_{n-1}(x), I_n(x)$ in \eqref{3.8a} and equating coefficients of $x^N$ gives
$
d_{n-1,N} = {1 \over n } ( B_n + N) d_{n,N} $.
Carrying out the analogous procedure in \eqref{6.1b} with $p=n-j$, $j=1,\dots,n-1$ gives
\[
d_{n-j-1,N} = {1 \over n - j } ( (B_{n-j} + N) d_{n-j,N} - j d_{n-j+1,N-1} ) .
\]
After successive substitutions, at the completion of step $j$, one has an expression for $d_{n-j-1,N}$ in terms of $\{d_{n,N'} \}_{N'=N-1-[j/2],\dots,N}$.
In relation to $d_{0,N}$ and $d_{1,N}$ there is the additional
equation~$
(N+1) d_{0,N+1} - n d_{1,N} = 0
$
as follows from \eqref{op3a}. Substituting the two previously obtained expressions for $d_{0,N}$, $d_{1,N}$ in terms of $\{d_{n,N'} \}$, we deduce \eqref{1.2Y}.
\end{proof}

As noted in the first line of this subsection, in the present context, $n=l, B_p = p (l-p)$. Further, with $\{T_l (N) \}$ as
in \eqref{1.2Y} we have $d_{l,N} = T_l (N) /(N!)^2$. In terms of $\{T_l(N) \}$, in low-order cases the procedure of the proof of Proposition~\ref{P3.4e} gives
\begin{gather}
(N+2)T_{2}(N+1) = (4N+2) T_{2}(N) , \nonumber\\
(N+3)^2 T_{3}(N+1) = \bigl(9+22 N + 10 N^2\bigr) T_{3}(N) - 9 N^2 T_{3}(N-1).\label{op4}
\end{gather}
Note that the solution of the first recurrence with $T_2(0)=1$ is
the Catalan number $T_2(N) = {1 \over N+1} \binom{2N}{N}$; see
\cite{St07} for references.
The procedure reclaims \eqref{1.2X} for $l=4$, and for $l=5$ gives the difference equation \eqref{1.2Y} with upper terminal on the sum $k=3$ and\footnote{{This corrects the corresponding recurrence in \cite[two displayed equations below (18)]{St07}, which is reported with~$P_3(N)$ of opposite sign.}}
\begin{align}
&P_0(N) = (N+4)^2(N+6)^2,\qquad P_1(N) = -375+400 N + 843 N^2 + 322 N^3 + 35 N^4, \nonumber \\
&P_2(N) = \bigl(45+622 N + 259 N^2\bigr)(N-1)^2,\qquad P_3(N) = - 255 (N-2)^2 (N-1)^2.\label{op5}
\end{align}
From the explicit forms \eqref{op4}, \eqref{1.2X} and \eqref{op5}, one sees an obvious pattern in that the degree of the polynomial coefficients within each case is the same, and equal to $l-1$. However, the cancellations which take place in going from the difference equation in Proposition~\ref{P3.4e} to the one for
$\{T_l(N) \}$ makes this difficult to anticipate.

\begin{Remark}\label{R4.1} \quad
\begin{itemize}\itemsep=0pt
\item[(1)] We know that the matrix differential equation of
Proposition~\ref{P3.2} has a unique solution with
$d_{p,0} = \delta_{p,l} (p=0,\dots,l)$ in \eqref{3.10a};
recall Remark \ref{R3.1} point~1. As a consequence, it must be that the scalar differential equation \eqref{MD3a2} has a unique power series solution subject only the normalisation $I_n(0) = 1$
(i.e.,~that the constant term in the power series is fixed). For this latter property to hold, it must be that the corresponding difference equation~\eqref{1.2Y} (corrected according to
Proposition~\ref{P3.4d}) must have the property that the coefficients of~${T_l(-1), T_l(-2),\dots, T_l(-[(l-1)/2])}$, whenever they appear, must vanish. Note that this property is illustrated in \eqref{1.2X} and \eqref{op5}.

\item[(2)] In relation to computing the power series solution {for general $l$}, there is no advantage in first reducing the matrix differential equation down to a scalar differential equation, rather use of the analogue of \eqref{MD3a2a} (which is based on the matrix differential equation) is more efficient. As already mentioned in Remark \ref{R3.1}, point 2, a quantification of the complexity when using this approach is given in Appendix \ref{appB}.
\end{itemize}
\end{Remark}

\subsection{Characterisation of the matrix integral in (\ref{1.6})}\label{S4.3}
The result of Proposition~\ref{P3.2} with
$\beta =2$, $n=l$, $B_p = p (2l-p)$ gives,
according to \eqref{2.8a} and \eqref{MDa2},
the matrix integral in \eqref{1.6} as proportional to $I_n(x)$.
 Following the method of Proposition~\ref{P3.4}, we can similarly specify the polynomials in the scalar linear differential equation reduction of the matrix differential equation from Proposition~\ref{P3.2} for small values of $l$ (without this latter restriction, the coefficients in the corresponding polynomials soon become unwieldy due to the number of digits involved and their increasing degree; recall the listing of
Proposition~\ref{P3.4}).

\begin{Proposition}\label{P3.5}
Consider the matrix integral in \eqref{1.6} with $z = x^{1/2}$, multiplied by \smash{$x^{-l^2/2}$} and up to a sign factor \smash{$(-1)^{\frac{l(l-1)}{2}}$}.
For $l=2,\dots,5$ this quantity satisfies the linear differential equations
\begin{gather*}
 \left (
 x^2 {{\rm d}^3 \over{\rm d} x^3} + 10 x {{\rm d}^2 \over{\rm d} x^2} + (20 - 4 x) {{\rm d} \over{\rm d} x} - 10 \right ) y(x) = 0, \\
 \left (
x^3 {{\rm d}^4 \over{\rm d} x^4} + 28 x^2 {{\rm d}^3 \over{\rm d} x^3} + (230 - 10 x) x {{\rm d}^2 \over{\rm d} x^2}+
(540-122 x) {{\rm d} \over{\rm d} x} + (-270 + 9x) \right ) y(x) = 0, \\
\left (
x^4 {{\rm d}^5 \over{\rm d} x^5} + 60 x^3 {{\rm d}^4 \over{\rm d} x^4} + (1238 - 20 x) x^2 {{\rm d}^3 \over{\rm d} x^3}+ (10268 - 662 x) x {{\rm d}^2 \over{\rm d} x^2}
\right.\\
\left.\qquad +\bigl(28288-6108 x +64 x^2\bigr) {{\rm d} \over{\rm d} x} + (-14144 +640 x) \right ) y(x) = 0 ,\\
\left (
x^5 {{\rm d}^6 \over{\rm d} x^6} + 110 x^4 {{\rm d}^5 \over{\rm d} x^5} + (4538 - 35 x) x^3 {{\rm d}^4 \over{\rm d} x^4}+ (87008 - 2422 x) x^2 {{\rm d}^3 \over{\rm d} x^3}
\right.\\
\qquad+\bigl( 765908 - 55800 x + 259 x^2\bigr) x{{\rm d}^2 \over{\rm d} x^2} +
\bigl( 2431000 -485408 x + 8392 x^2\bigr)
{{\rm d} \over{\rm d} x} \\
\qquad +
\left.\bigl(-1215500 + 59300 x -225 x^2 \bigr) \right ) y(x) = 0,
\end{gather*}
respectively.
\end{Proposition}

As revised in Appendix~\ref{appendixA}, the work \cite{KW24b} has given an alternative way to derive the degree~${l+1}$ scalar linear differential equation satisfied by
the matrix integral \eqref{1.6}. In the case $l=2$, this is made explicit in
\cite[equation~(97)]{KW24b}, and consistency with the third-order equation listed above can be checked. Implementing the recursive strategy leading to
\eqref{generalhigherorderequation} of Appendix~\ref{appendixA}, in the cases $l=3,4,5$ using computer algebra establishes consistency of the results obtained by the alternative method for all the differential equations listed in Proposition~\ref{P3.5}.

Motivating the investigation of \cite{KW24b} in this direction was the realisation that the formalism given in \cite{FW06} relating to a characterisation in terms of a Painlev\'e III$'$ $\sigma$-form transcendent cannot be used to generate the full power series about the origin due to the stated boundary condition therein not uniquely determining the coefficients.
Thus,
considered in \cite{FW06} was the quantity
\[
 v(s) := - s {{\rm d} \over {\rm d}s} \log \big ( s^{l^2/2}
 {\rm e}^{-s/4} \big \langle
 (\det U)^l
 {\rm e}^{{1 \over 2} \sqrt{s} \operatorname{Tr}(U + U^\dagger)} \big \rangle_{U(l)} \big );
 \]
 cf.~\eqref{s1}. Using results from \cite{FW02} it was noted that $v(s)$ satisfies the particular $\sigma$-Painlev\'e III$'$ equation
 \[
(sv'')^2 + v' (4 v' - 1)(v - sv') - {l^2 \over 16} = 0,
\]
(cf.~\eqref{us})
subject to the small-$s$ boundary condition
\begin{equation}\label{us1x}
v(s) = - l^2 + {s \over 8} + {\rm O}\bigl(s^{2}\bigr).
\end{equation}
It was shown in \cite{FW06} that this allowed for the generation of the first $l$ even powers recursively, as was the need of the application therein. However, missed in \cite{FW06} was the fact made explicit in~\cite{KW24b} that the boundary condition \eqref{us1x} does not uniquely determine a solution of \eqref{us1x}, with there in fact being a family of solutions which require the specification of the coefficient of $s^{2l+1}$ to be made unique beyond degree $s^{2l}$, and which was incorrectly set to zero {(albeit without consequence to the application given there related to the moments of the first derivative of the Riemann zeta function)}.

The finding in \cite{KW24b} of a scalar differential equation characterisation of
the matrix integral \eqref{1.6} overcame this latter point, and was used to obtain in particular the coefficients up to degree~$s^{4l}$ as required for the application to the moments of the second derivative of the Riemann zeta function on the critical line as outlined in the text of the Introduction including \eqref{1.8}. {The present work complements such a computational objective, with the vector recurrence \eqref{MD3a0} as deduced from the matrix differential equation and analysed from a complexity viewpoint in the second subsection of Appendix~\ref{appB}, providing an efficient scheme for this purpose}.

\appendix

\section[Alternative derivations from Keating and Wei (2024)]{Alternative derivations from \cite{KW24b}}\label{appendixA}

The purpose of this appendix is to outline the method given in \cite{KW24b} for the derivation of the degree $l+1$ linear differential equation satisfied by~\eqref{1.6} (see Corollary \ref{altermethod} below), which at a~practical level allows for an alternative derivation of the results of Proposition~\ref{P3.5}. Moreover, the working of \cite{KW24b} will be extended to allow for consideration of the family of matrix averages in \eqref{2.8a}
with $q=l+\alpha$ (for $q \in \mathbb Z_{\ge 0}$), or more generally
the family of Hankel determinants in~\eqref{1.6} with the index of the $I$-Bessel function therein generalised from $j+k+1$ to $j+k+\alpha$ (for~${\alpha \in \mathbb R_{\ge 0}}$).
The starting point of the method is the connection between a shifted version of the Hankel determinant appearing in \eqref{1.6}
\begin{equation}\label{defoftaulz}
\tau_{l}(x) := \det\!\big[I_{j+k+1}\bigl(2\sqrt{x}\bigr)\big]_{j,k=0}^{ l-1},
\end{equation}
and the derivative of the Hankel determinant itself. Specifically, we begin with the definition of the determinant of Hankel matrices whose column indices are shifted by Young diagrams.
\begin{Definition}
Let $\alpha\in \mathbb{R}_{\geq 0}$. Let $l\geq 1,h\geq 0$ be integers. Let $Y=(m_{1},\ldots,m_{h})$ be a~Young diagram. Define
\smash{$
\tau_{l,Y,\alpha}(x):=\det\!\big[I_{j+k+\alpha+m_{l-k}}\bigl(2\sqrt{x}\bigr)\big]_{j,k=0}^{ l-1}$}.
Here we adopt the convention that~$m_{h+1}=\cdots=m_{l}=0$ whenever $h<l$. When $Y=\varnothing$, we simply write $\tau_{l,\varnothing,\alpha}$ as $\tau_{l,\alpha}$ for short.
\end{Definition}
In particular, when $Y=\varnothing$ and $\alpha=1$, the above definition reduces to $\tau_{l}(x)$ as given in \eqref{defoftaulz}.
Using the trace of a square matrix, we define the following operations.
\begin{Definition}
Let $\alpha\in \mathbb{R}_{\geq 0}$. Let $l\geq 1$, $h,n\geq 0$ be integers. Let $Y=(m_{1},\ldots,m_{h})$ be a~Young diagram. Define
\begin{equation*}
(T_{n}\tau_{l,Y,\alpha})(x):=\operatorname{Tr}\left(\mathrm{adj}\bigl(\big[I_{j+k+\alpha+m_{l-k}}(2\sqrt{x}\bigr)\big]_{j,k=0}^{ l-1})\big[I_{j+k+\alpha+m_{l-k}+n}\bigl(2\sqrt{x}\bigr) \big]_{j,k=0}^{ l-1}\right)
\end{equation*}
and
\begin{align*}
(S_{n}\tau_{l,Y,\alpha})(x):={}&\operatorname{Tr}\bigl(\mathrm{adj}\bigl(\big[I_{j+k+\alpha+m_{l-k}}\bigl(2\sqrt{x}\bigr) \big]_{j,k=0}^{ l-1}\bigr)\\
&\times\big[(j+k+\alpha+m_{l-k}+n)I_{j+k+\alpha+m_{l-k}+n}\bigl(2\sqrt{x}\bigr) \big]_{j,k=0}^{ l-1}\bigr),
\end{align*}
where for a matrix $M$, $\mathrm{adj}(M)$ denotes its adjugate.
\end{Definition}
In deriving the higher-order linear differential equation for $\tau_{l,\alpha}$, we mainly use a special form of Young diagram, namely
\begin{equation}\label{defofhookdiagram}
Y_{m,j} := (m-j+1,1,1,\ldots,1),
\end{equation}
of weight $m$, where $m \geq j\geq 1$.
By applying \cite[Theorem~3.1]{KW24b} to general Hankel determinants shifted by $Y_{m,j}$ and specializing to $\tau_{l,Y_{m,j},\alpha}$, we obtain the following inverse linear system.

\begin{Proposition}
\label{gerenel linear systems}
Let $l,m\geq 1, m\geq j\geq 1$. Let $Y_{m,j}$ be given as in \eqref{defofhookdiagram}. Then
\begin{equation*}
\label{linear system 1}
\begin{pmatrix}
\tau_{l,Y_{m,1},\alpha} \\
\vdots \\
\vdots \\
\tau_{l,Y_{m,m},\alpha} \\
\end{pmatrix}
= \mathbf B_{m}
\begin{pmatrix}
\vdots \\
\sum_{h=1}^{j-1} (-1)^h T_{h} \tau_{l,Y_{m-h,j-h},\alpha} \\
\vdots \\
T_{m}\tau_{l,\alpha} \\
\end{pmatrix}_{j=2,\ldots,m},
\end{equation*}
where
$\mathbf B_{m}=(b_{j,k})_{j,k=1,\ldots,m}$ is the $m\times m$ matrix satisfying
\begin{equation}\label{definition of B}
b_{j,k} =\begin{cases}
(-1)^{j+k-1}/k(k+1), & {\color{black} j\leq k \leq m-1}, \\
-1/j, & k=j-1, \\
0 ,& k<j-1, \\
(-1)^{j-1}/m ,& k=m.
\end{cases}.
\end{equation}
\end{Proposition}
Using the recurrence relations of the modified Bessel function of the first kind,
\begin{gather}
\frac{\rm d}{{\rm d}x} I_{\beta}\bigl(2\sqrt{x}\bigr) =
\frac{I_{\beta+1}\bigl(2\sqrt{x}\bigr) }{\sqrt{x}} + \frac{\beta}{2x}I_{\beta}\bigl(2\sqrt{x}\bigr) , \nonumber\\
I_{\beta+2}\bigl(2\sqrt{x}\bigr) =I_{\beta}\bigl(2\sqrt{x}\bigr) -\frac{\beta +1}{\sqrt{x}}I_{\beta+1}\bigl(2\sqrt{x}\bigr) \label{recursiveformula2},
\end{gather}
we can express $\tau_{l,Y_{1,1},\alpha}$, $\tau_{l,Y_{2,1},\alpha}$, and $\tau_{l,Y_{2,2},\alpha}$ in terms of the derivatives of $\tau_{l,\alpha}$. The following lemmas for general $\alpha$ are obtained by a similar argument to that used in \cite[Propositions~4.1 and~4.2]{KW24b} for $\alpha=1$.
\begin{Lemma}\label{length1}
Let $l\geq 1,m\geq 0$ and $Y$ be a Young diagram of length $m$. Then
\begin{equation*}
T_{1}\tau_{l,Y,\alpha}(x) = \sqrt{x}\frac{\rm d}{{\rm d}x} \tau_{l,Y,\alpha}(x) - \frac{l^2+m+l(\alpha-1)}{2\sqrt{x}}\tau_{l,Y,\alpha}.
\end{equation*}
In particular, if $Y=\varnothing$, then
\begin{equation}\label{initial1}
\tau_{l,Y_{1,1},\alpha}(x) = \sqrt{x}\frac{\rm d}{{\rm d}x} \tau_{l,\alpha}(x) - \frac{l^2+l(\alpha-1)}{2\sqrt{x}}\tau_{l,\alpha} .
\end{equation}
\end{Lemma}
\begin{Lemma}\label{length2}
\label{initial values of tau}
Let $l\geq 1$.
Then
\begin{align}
\tau_{l,Y_{2,1},\alpha} ={}&\frac{x}{2} \frac{{\rm d}^2}{{\rm d}x^2} \tau_{l,\alpha} - \frac{l(l+\alpha+1)+\alpha-1}{2} \frac{\rm d}{{\rm d}x} \tau_{l,\alpha} \nonumber\\
&+ \frac{\bigl(l^2+l(\alpha-1)\bigr)\bigl(l^2+l(\alpha+3)+2\alpha\bigr)+4lx}{8x} \tau_{l,\alpha},\label{initial21} \\
\tau_{l,Y_{2,2},\alpha} ={}& \frac{x}{2} \frac{{\rm d}^2}{{\rm d}x^2} \tau_{l,\alpha} - \frac{l(l+\alpha-3)-(\alpha-1)}{2} \frac{\rm d}{{\rm d}x} \tau_{l,\alpha}\nonumber\\
&+ \frac{\bigl(l^2+l(\alpha-1)\bigr)\bigl(l^2+l(\alpha-5)+4-2\alpha\bigr)-4lx}{8x} \tau_{l,\alpha}\label{initial22}.
\end{align}
\end{Lemma}
For a general weight $m$ of Young diagrams, we can recursively obtain an expression for $\tau_{l,Y_{m,j},\alpha}$ in terms of the derivatives of $\tau_{l,\alpha}$.
Thus,
by identity \eqref{recursiveformula2}, we can rewrite operations involving~$T_{h}$ in Proposition~\ref{gerenel linear systems} in terms of $S_{h-1}$ and $T_{h-2}$. Combining Lemmas \ref{length1} and \ref{length2}, we then obtain the following formula for $\tau_{l,Y_{m,j},\alpha}$, which depends recursively on $m$.
\begin{Proposition}
\label{recursive formula for tau0}
Let $l\geq 1$ and $m\geq 3$. Let $m\geq j\geq 1$. Let $Y_{m,j}$ be given as in \eqref{defofhookdiagram}.
Then
\begin{align}
\begin{pmatrix}
\tau_{l,Y_{m,1},\alpha} \\
\vdots \\
\tau_{l,Y_{m,m},\alpha}
\end{pmatrix}
={}&
- \sqrt{x} \mathbf B_{m}
\begin{pmatrix}
\frac{\rm d}{{\rm d}x} \tau_{l,Y_{m-1,1},\alpha} \\
\vdots \\
\frac{\rm d}{{\rm d}x}\tau_{l,Y_{m-1,m-1},\alpha} \\
0
\end{pmatrix}
+\frac{l^2+l(\alpha-1)+m-1}{2\sqrt{x}} \mathbf B_{m}\begin{pmatrix}
 \tau_{l,Y_{m-1,1},\alpha} \\
\vdots \\
\tau_{l,Y_{m-1,m-1},\alpha} \\
0
\end{pmatrix} \nonumber\\
& - \frac{1}{\sqrt{x}} \mathbf C^{(1)}_{m}
\begin{pmatrix}
 \tau_{l,Y_{m-1,1},\alpha} \\
\vdots \\
\tau_{l,Y_{m-1,m-1},\alpha}
\end{pmatrix}
+ \mathbf C^{(2)}_{m}
\begin{pmatrix}
 \tau_{l,Y_{m-2,1},\alpha} \\
\vdots \\
\tau_{l,Y_{m-2,m-2},\alpha}
\end{pmatrix},\label{recursiveprocess}
\end{align}
where $\mathbf B_{m}$ is given as in \eqref{definition of B}, \smash{$\mathbf C^{(1)}_{m}= \big(c_{j,k}^{(1)}\big)_{\substack{j=1,\ldots,m
\\k=1,\ldots,m-1}}$} {is} an $m \times (m-1)$ matrix satisfying
\[
c_{j,k}^{(1)}=
\begin{cases}
(-1)^{j+k}\bigl(\frac{m-1-k}{k+1}+\frac{2l-k+\alpha}{k(k+1)}\bigr) ,& \text{if } j\leq k\leq m-1,\\
 -\frac{1}{k+1}(k(2l-k+\alpha)+m-1-k),& \text{if } k=j-1, \\
 0, & \text{if } k<j-1,
\end{cases}
\] and
\smash{$\mathbf C^{(2)}_{m}= \bigl(c_{j,k}^{(2)}\bigr)_{\substack{j=1,\ldots,m \\k=1,\ldots,m-2}}$} is an $m\times (m-2)$ matrix satisfying
\[
c_{j,k}^{(2)} =
\begin{cases}
 (-1)^{j+k} \frac{l+2}{(k+1)(k+2)}, & j-1\leq k \leq m-2, \\
\frac{j-l-2}{j}, & k=j-2, \\
 0 ,& k<j-2.
\end{cases}
\]
\end{Proposition}
\begin{Remark}
The above proposition generalizes \cite[Proposition 4.3]{KW24b} from the case $\alpha=1$ to general $\alpha$.
It can be rewritten as a first-order matrix linear differential equation
for the vector on the left-hand side analogous to (but distinct from) that implied by Proposition~\ref{P3.2}, with $n=m$, $\beta =2$
and $B_p$ given by \eqref{MD3a3} with $q=m+\alpha$. On this latter point, note that the entries in the corresponding vectors agree in the final entry but involve different functions otherwise.
\end{Remark}

\begin{Theorem}\label{higherorderequationdorgeneralparameter}
Let $l \geq 1$. Suppose that $\tau_{l,Y_{l+1,l+1},\alpha}$ is obtained recursively by formula \eqref{recursiveprocess} with initial conditions \eqref{initial1}, \eqref{initial21}, and \eqref{initial22}. Then
\begin{gather}
\label{taukYss}
\tau_{l,Y_{l+1,l+1},\alpha}(x)
= x^{-\frac{l+1}{2}} \sum_{n=0}^{l+1} x^n
\left(\sum_{j=0}^{[\frac{l+1-n}{2}]} b_{j,n}^{(l+1)}(\alpha) x^j \right)
\frac{{\rm d}^n \tau_{l,\alpha}(x)}{{\rm d}x^n},
\end{gather}
where the coefficients \smash{$b_{j,n}^{(l+1)}(\alpha)$} are given recursively. Moreover, we have the following equation:
\begin{gather}\label{generalhigherorderequation}
\sum_{n=0}^{l+1} x^n
\left(\sum_{j=0}^{[\frac{l+1-n}{2}]} b_{j,n}^{(l+1)}(\alpha) x^j \right)
\frac{{\rm d}^n \tau_{l,\alpha}(x)}{{\rm d}x^n} = 0.
\end{gather}
\end{Theorem}
\begin{proof}
By induction, it is straightforward to verify that \eqref{taukYss} holds. It now remains to prove~\eqref{generalhigherorderequation}. On one hand, by Proposition~\ref{recursive formula for tau0},
\begin{align}
\tau_{l,Y_{l+1,l+1},\alpha}(x) ={}& \frac{1}{l+1} \left(\sqrt{x} \frac{{\rm d}\tau_{l,Y_{l,l},\alpha}(x)}{{\rm d}x} - \frac{l^2+l\alpha}{2\sqrt{x}}\tau_{l,Y_{l,l},\alpha}(x) \right) + \frac{l(l+\alpha)}{(l+1)\sqrt{x}} \tau_{l,Y_{l,l},\alpha}(x)\nonumber\\
&-\frac{1}{l+1} \tau_{l,Y_{l-1,l-1},\alpha}(x)\label{from therecursiveformula} .
\end{align}
On the other hand, note that
\begin{equation*}
\tau_{l,Y_{l-1,l-1},\alpha}(x)=\operatorname{Tr}\bigl(\mathrm{adj}\bigl(\big[I_{j+k+\alpha+1}\bigl(2\sqrt{x}\bigr)\big]_{j,k=0}^{ l-1}\bigr)\big[I_{j+k+\alpha}\bigl(2\sqrt{x}\bigr) \big]_{j,k=0}^{ l-1}\bigr).
\end{equation*}
By the observation \smash{$\tau_{l,Y_{l,l},\alpha}(x)= \det \bigl(I_{j+k+\alpha+1}\bigl(2\sqrt{x}\bigr) \bigr)_{j,k=0,\ldots,l-1}$}
and the recurrence relation
\begin{equation*}
I_{j+k+\alpha}\bigl(2\sqrt{x}\bigr)
=\sqrt{x} \frac{\rm d}{{\rm d}x} I_{j+k+\alpha+1}\bigl(2\sqrt{x}\bigr) + \frac{\alpha+j+k+1}{2\sqrt{x}} I_{j+k+\alpha+1}\bigl(2\sqrt{x}\bigr) ,
\end{equation*}
we have
\begin{equation*}
\sqrt{x} \frac{\rm d}{{\rm d}x} \tau_{l,Y_{l,l},\alpha}(x) + \frac{l^2+l\alpha}{2\sqrt{x}} \tau_{l,Y_{l,l},\alpha}(x)
-\tau_{l,Y_{l-1,l-1},\alpha}(x) = 0.
\end{equation*}
Then by \eqref{from therecursiveformula}, we have the identity
$
\tau_{l,Y_{l+1,l+1},\alpha}\equiv 0$.
Hence, using above and \eqref{taukYss}, we obtain~\eqref{generalhigherorderequation}.
\end{proof}
\begin{Remark}
Not only $\tau_{l,Y_{l+1,l+1},\alpha}$, but also $\tau_{l,Y_{m,j},\alpha}$ for any $m,j$ with $l+1 \leq j \leq m$, which can be obtained recursively from formula \eqref{recursiveprocess} as explained in \cite[Remark~3.5]{KW24b}, should be $0$. Hence, this will also lead to a linear differential equation of a form similar to \eqref{generalhigherorderequation}, but of degree $m$.
\end{Remark}
As a consequence of Theorem \ref{higherorderequationdorgeneralparameter} with $\alpha=1$, we obtain the following result, which provides a~method to derive the degree $l+1$ linear differential equation satisfied by \eqref{1.6}.
\begin{Corollary}\label{altermethod}
Let $l\geq 1$. Let \smash{$b_{j,n}^{(l+1)}(1)$} be given as in Theorem {\rm\ref{higherorderequationdorgeneralparameter}} with $\alpha=1$. Then
\begin{equation*}
\sum_{n=0}^{l+1} x^n
\left(\sum_{j=0}^{[\frac{l+1-n}{2}]} b_{j,n}^{(l+1)}(1) x^j \right)
\frac{{\rm d}^n \tau_{l}(x)}{{\rm d}x^n} = 0.
\end{equation*}
\end{Corollary}
Finally, applying Corollary \ref{altermethod} and verifying with the code,\footnote{\url{https://drive.google.com/file/d/1gd4kQO9ABKhALrzeYos-DbeLcelQmik6/view?usp=share_link}} we re-obtain Proposition~\ref{P3.5}.

\renewcommand{\thefootnote}{$*$}

\section[Computational aspects. Appendix by Folkmar Bornemann]{Computational aspects. Appendix by Folkmar Bornemann\footnote{Department of Mathematics, Technical University of Munich, 80290 M\"unchen, Germany}\renewcommand{\thefootnote}{}
\footnote{E-mail: \href{mailto:bornemann@tum.de}{bornemann@tum.de}}}\label{appB}

\subsection[D-finiteness]{$\boldsymbol{D}$-finiteness}\label{appendixB1}

\newcommand{\C}{{\mathbb C}}
\newcommand{\Q}{{\mathbb Q}}
\newcommand{\Z}{{\mathbb Z}}

\renewcommand{\thefootnote}{\arabic{footnote}}
\setcounter{footnote}{5}

We discuss the unitary matrix integrals \eqref{1.3} and \eqref{1.6} given in the form
\[
y_l(z) := \det\left[ f_{j-k}(z)\right]_{j,k=0,\ldots,l-1},\qquad
w_l(z) := c_l(-1)^{l(l-1)/2}z^{-l^2} \det\left[ f_{j+k+1}(z)\right]_{j,k=0,\ldots,l-1},
\]
which are Toeplitz and Hankel determinants of
$
f_{n}(z) := I_n(2z)$,
where $I_n$ denotes the modified~Bessel function of the first kind of integer order $n$. By the symmetries of the unitary matrix integrals, $y_l$ and $w_l$ are {\em even} entire functions. By choosing the normalization $c_l = G(2l+1)/G(l+1)^2$ (where $G$ is the Barnes $G$-function), their power series start as\footnote{{The normalization of $w_l$ follows from \eqref{2.8a} by noting that $_0^{\mathstrut} F_1^{(1)}\bigl(2l; \bigl(z^2\bigr)^l\bigr) = 1 + \frac{z^2}{2} + \cdots.$}}
\[
y_l(z) = 1 + z^2 + \cdots, \qquad w_l(z) = 1 + \frac{z^2}{2} + \cdots.
\]

Bessel's differential equation
$
z^2 f_n'' + z f_n' - \bigl(4z^2 + n^2\bigr)f_n=0
$
shows that the $f_n$ are $D$-finite (holonomic), that is, they satisfy a linear differential equation with polynomial coefficients. Since $D$-finite power series constitute a subalgebra of the formal power series $\C\llbracket z\rrbracket$ (see \cite[Theorem~2.3]{St80}), it is immediately clear that $y_l$, $w_l$ are also $D$-finite. A symmetry argument shows that the holonomic equations for the even functions $y_l$ and $w_l$ transform to holonomic equations of the same order for $Y_l(x):=y_l\bigl(\sqrt{x}\bigr)$ and $W_l(x):=w_l\bigl(\sqrt{x}\bigr)$ -- so these entire functions are also $D$-finite.
$D$-finite power series have $P$-recursive (holonomic) coefficients (and vice versa, see~\cite[Theorem~1.5]{St80}), that is, the coefficients satisfy a linear recurrence equation with polynomial coefficients. Besides being of structural interest, such a recurrence allows for the effective calculation of the power series coefficients.
Though the general theory of holonomic systems provides simple bounds on the order of the differential equations and recurrences -- so that they can be constructed by Groebner bases elimination methods, see \cite[Section 4]{Z} -- those bounds are often too large to be of practical use. Since the Bessel differential equation is of order~$2$, the degree of the holonomic differential equation for~$y_l$, or $w_l$, is superficially bounded by $2^l l!$ because the determinants are
 sums of $l!$ terms, each of which is the product of $l$ such Bessel functions.

However, by identifying the family $f_n(z)$ as a discrete-continuous holonomic system in $(n,z)$ (cf.\ \cite[p.~322]{Z}), the order bound can be significantly reduced to the optimal $l+1$ (which is the order given in Proposition~\ref{P3.3}) and the coefficients of the higher-order linear differential equation can be computed by linear algebra.
To this end, we supplement the Bessel differential equation with the holonomic relations between $f_{n+1}$ and $f_n$ (cf.\ \eqref{recursiveformula2}), namely
\[
2 z f_{n+1} = z f_n' - n f_n,\qquad 2z^2 f_{n+1}' = -(n+1) z f_n' + \bigl(4z^2 + n(n+1)\bigr)f_n.
\]
Inserting this into the determinants, we get
\[
y_l, w_l \in V_l := \text{span}_{\C(z)}\big\{ (f_0)^j (f_0')^{l-j}\colon j=0,\ldots,l\big\};
\]
clearly we have $\dim_{\C(z)} V_l \leq l+1$. A classical result of Siegel \cite[Satz 9]{S29}\footnote{{Since the Galois group of the differential equation for the Bessel function $J_\nu$ ($\nu\not\in \frac{1}{2} + \Z$) is $\text{SL}_2(\C)$ (cf.~\cite[Appendix]{K68}), any fundamental system and their derivatives have transcendence degree $\dim_{\C} \text{SL}_2(\C) = 3$ over $\C(z)$.}} shows that $f_0$, $f_0'$ are algebraically independent over $\C(z)$, so that $\dim_{\C(z)} V_l = l+1$ and $z$, $f_0$, $f_0'$ can be treated as three independent indeterminates: calculations take place in the polynomial ring $\C[z,f_0,f_0']$. Since Bessel's equation shows that $V_l$ is closed under differentiation, there is a {\em unique} set of relatively prime polynomials $a_j \in \C[z]$, normalized to monic $a_{l+1}$, such that
\begin{equation}\label{eq:deq}
a_{l+1} y_l^{(l+1)} + a_l y_l^{(l)} + \cdots + a_1 y_l' + a_0 y_l = 0;
\end{equation}
the same holds for $w_l$. The $a_j$ are computed by solving the homogeneous linear system of dimension $(l+1)\times (l+2)$ over
$\C(z)$ of full rank that is obtained by setting the coefficients of the monomials $f_0^j (f_0')^{l-j}$ on the left of \eqref{eq:deq} to zero. Using a computer algebra system, it is thus straightforward to reclaim the differential equations tabulated in Propositions \ref{P3.4} and \ref{P3.5}.

\begin{Remark} In \cite{BG00}, using the order bound $l+1$, the differential equations, such as \eqref{1.5} for~${l=4}$, were obtained in a far less efficient way by first tabulating, using other means, the first~$N$ entries of the power series of $Y_l(x) = y_l\bigl(\sqrt{x}\bigr)$
for sufficiently large $N$.
Using degree bounds for the coefficient polynomials $a_j$, the presumed differential equation generates a set of linear equations for the coefficients of $a_j$ (normalized to monic $a_{l+1}$) being compatible with the tabulated values. If $N$ is large enough, and the bounds are chosen appropriately, the linear system has a unique solution. This procedure has been implemented in the Maple package {\tt gfun}~\cite{SZ94}; as an example, the differential equation \eqref{1.5} is obtained by the command {\tt listtodiffeq} after adding three more terms to the list \eqref{1.1} as follows:

\begin{minipage}{0.8\textwidth}{\small
\begin{verbatim}
with(gfun):
l := 4:
L := [1, 1, 2, 6, 24, 119, 694, 4582, 33324, 261808, 2190688, 19318688, 178108704,
		    1705985883, 16891621166, 172188608886]:
c := [seq(L[k + 1]/(k!)^2, k = 0 .. 15)]:
gfun:-Parameters(`maxordereqn' = l + 1, `maxdegcoeff' = l):
listtodiffeq(c, Y(x));
\end{verbatim}}
\end{minipage}
\medskip

\noindent The recursion \eqref{1.2X} is thereafter obtained by executing the {\tt gfun}-command {\tt listtorec} as follows:

\medskip
\begin{minipage}{0.8\textwidth}{\small
\begin{verbatim}
listtorec(L, T(N));
\end{verbatim}}
\end{minipage}
\end{Remark}

\subsection[Complexity of tabulating T\_l(N)]{Complexity of tabulating $\boldsymbol{T_l(N)}$}\label{appB2}

As already noted, $T_l(N)=N!$ for $N\leq l$, so compiling a table of those numbers of combinatorial interest can be restricted to the triangular array $1\leq l\leq N\leq N_*$. Here we compare a tabulation as based on \eqref{MD3a0}, to that used in the recent works \cite{Bo24,FM23}.\footnote{{For a history of tabulation by combinatorial methods, see \cite[p.\ 916]{Bo24}. A tabulation for $N_*=1000$, computed with the method in \cite[Section~3.2]{Bo24}, is available for download at \url{https://box-m3.ma.tum.de/f/7c4f8cb22f5d425f8cff/}.}}
There, instead of considering \eqref{1.0} directly, one first aims to provide a power series expansion of what is essentially the logarithmic derivative
 \big(with $z = {1 \over 2} \sqrt{s}$\big),
 \begin{equation}\label{s1}
 u(s) := - s {{\rm d} \over {\rm d}s} \log \big (
 {\rm e}^{-s/4} \big \langle {\rm e}^{{1 \over 2} \sqrt{s} \operatorname{Tr}(U + U^\dagger)} \big \rangle_{U(l)} \big ).
 \end{equation}
 The significance of \eqref{s1} is the characterisation of $u(s)$ as the solution of the second-order second-degree $\sigma$-Painlev\'e III$'$ equation \cite{FW02,TW94b}
\begin{equation}\label{us}
\bigl(su''\bigr)^2 -\bigl(l u'\bigr)^2 +u' \bigl(4 u' - 1\bigr)\bigl(u - su'\bigr) = 0
\end{equation}
with boundary condition for small $s$
\begin{equation}\label{us1}
u(s) = {s^{l+1} \over 4^{l+1} l!(l+1)!} + {\rm O}\bigl(s^{l+2}\bigr).
\end{equation}
Writing $u(s)$ as a power series consistent with \eqref{us1},
\[
u(s) = \sum_{n=l+1}^\infty c_n s^n
\]
 and substituting in \eqref{us} provides a recurrence for successive coefficients. However, since the differential equation
 \eqref{us} is {\em cubic} as a polynomial expression of $u$, $u'$, $u''$, the recursion is cubic in the coefficients $c_n$, involving double sums for the evaluation of the Cauchy products. It was therefore suggested in \cite[Section~3.2]{Bo24}, to use the equivalent third-order first-degree Chazy-I equation, which is obtained from differentiating \eqref{us} and dividing the result by $2u''$
\begin{equation}\label{us-ChazyI}
s^2u''' + s u'' -6s u'^2 +4 u u' +\bigl(s-l^2\bigr)u' -\frac{1}{2}u = 0.
\end{equation}
The differential equation is now {\em quadratic} as a polynomial expression in $u$, $u'$, $u''$, $u'''$ which yields a much simpler recursive formula for the $c_n$, $n=l+1,\ldots$ -- quadratic in the $c_n$ with just a single sum for the evaluation of the Cauchy products
\[
(n+1)\bigl(n^2-l^2\bigr)c_{n+1} + (n-\tfrac12)c_n - 2\sum_{m=l+1}^{n-l} m c_m \cdot (3(n-m)+1) c_{n+1-m} = 0.
\]
In this way, after integrating and exponentiating the truncated power series of $u(s)$ as required by \eqref{s1}, we get the initial coefficients $T_l(N)$ in the power series expansion of the average in~\eqref{1.2}.

Counting the operations in $\Q$, and noting that the fast exponentiation of truncated power series with $N_*$ terms takes just $O(N_*\log N_*)$ operations with a small implied constant (a method originating with R. Brent, see \cite{BS09}), computing a single row with a fixed $l$ accounts for
\[
O\bigl((N_*-l)^2\bigr) + O(N_*\log N_*)
\]
operations. From this, we get the first row in Table \ref{tab:1}.\footnote{{Note that the total cost when using the $\sigma$-Painlev\'e III$'$ equation as in \cite{FM23} is $O\bigl(N_*^4\bigr)$.}}

\begin{table}[tbp]
\caption{Cost for computing $T_l(N)$, $1 \leq l\leq N \leq N_*$; single row means fixed $l$; $\theta$, $c$ in the last two columns are subject to $0 < \theta$, $c < 1$.}
\label{tab:1}
\medskip
\centering
{\footnotesize
\begin{tabular}{p{2.4cm}p{1.4cm}p{4.0cm}p{3cm}p{2.8cm}}
\hline\noalign{\smallskip}
method based on & total cost & single row $l$ & single row $l = O(N_*^\theta)$ & single row $l = c N_*$ \\
\noalign{\smallskip}\Xhline{2\arrayrulewidth}\noalign{\smallskip}
Chazy-I \eqref{us-ChazyI} & $O\bigl(N_*^3\bigr)$ & $O\bigl((N_* - l)^2\bigr) + O(N_* \log N_*)$ & $O\bigl(N_*^2\bigr)$ & $(1-c)^2 \cdot O\bigl(N_*^2\bigr)$\\\noalign{\smallskip}
recursion \eqref{MD3a0} & $O\bigl(N_*^3\bigr)$ & $ O(l\cdot N_*)$ & $O\bigl(N_*^{1+\theta}\bigr)$ & $c\cdot O\bigl(N_*^2\bigr)$\\\noalign{}
\noalign{\smallskip}\hline
\end{tabular}}

\end{table}

On the other hand, the recurrence \eqref{MD3a0} of Proposition~\ref{P3.2} provides a novel alternative to the use of a nonlinear differential equation for purposes of computing the tabulation. In this formalism
$T_l(N^*)$ is read off from the vector $\mathbf d_{N^*}$. With the vector $\mathbf d_0$ known from \eqref{MDa2x}, successive vectors are computed by solving the lower triangular linear system in the recurrence~\eqref{MD3a0} by forward substitution. This only takes $O(l)$ operations since the $l\times l$ matrix on the right-hand side is bidiagonal. With $N^*$ iterations, the total number of steps is thus of the order $O(l \cdot N^*)$. From this, we get the second row in Table \ref{tab:1}.

So, whereas the Chazy-I based computation can take advantage of the shorter rows for $l \approx N_*$, a computation based on the recursion \eqref{MD3a0} benefits from the underlying comparatively small dimension $l$ for rows with $l \ll N_*$. Overall, when computing the full table, these differences are balanced, and both methods exhibit the same $O\bigl(N_*^3\bigr)$ complexity.

\subsection*{Acknowledgements}
The work of Peter J.~Forrester was supported by the Australian Research Council Discovery Project
DP250102552. The work of Fei Wei was supported by the Royal Society, grant URF$\backslash$R$\backslash$231028. Both authors thank the organisers of the MATRIX program ``Log-gases in Caeli Australi'', held in Creswick, Victoria, Australia during the first half of August 2025 for facilitating this collaboration, and for the event itself. We are grateful to the anonymous referees for their insightful comments and suggestions, which helped improve the clarity and presentation of this work.

\pdfbookmark[1]{References}{ref}
\LastPageEnding


\begin{thebibliography}{99}
\footnotesize\itemsep=0pt

\bibitem{A+24}
Alvarez E., Conrey J.B., Rubinstein M.O., Snaith N.C., Moments of the
 derivative of the characteristic polynomial of unitary matrices,
 \href{https://doi.org/10.1142/S2010326325500029}{\textit{Random Matrices
 Theory Appl.}} \textbf{14} (2025), 2550002, 45~pages,
 \href{http://arxiv.org/abs/2407.13124}{arXiv:2407.13124}.

\bibitem{Ao87}
Aomoto K., Jacobi polynomials associated with {S}elberg integrals,
 \href{https://doi.org/10.1137/0518042}{\textit{SIAM~J. Math. Anal.}}
 \textbf{18} (1987), 545--549.

\bibitem{ABGS21}
Assiotis T., Bedert B., Gunes M.A., Soor A., On a distinguished family of
 random variables and {P}ainlev\'e equations,
 \href{https://doi.org/10.2140/pmp.2021.2.613}{\textit{Probab. Math. Phys.}}
 \textbf{2} (2021), 613--642,
 \href{http://arxiv.org/abs/2009.04760}{arXiv:2009.04760}.

\bibitem{AGKW24}
Assiotis T., Keating J.P., Gunes M.A., Wei F., Exchangeable arrays and
 integrable systems for characteristic polynomials of random matrices,
 \href{http://arxiv.org/abs/2407.19233}{arXiv:2407.19233}.

\bibitem{A+25}
Assiotis T., Keating J.P., Gunes M.A., Wei F., Joint moments for characteristic
 polynomials from the orthogonal and symplectic groups,
 \href{http://arxiv.org/abs/2508.09910}{arXiv:2508.09910}.

\bibitem{AKW22}
Assiotis T., Keating J.P., Warren J., On the joint moments of the
 characteristic polynomials of random unitary matrices,
 \href{https://doi.org/10.1093/imrn/rnab336}{\textit{Int. Math. Res. Not.}}
 \textbf{2022} (2022), 14564--14603,
 \href{http://arxiv.org/abs/2005.13961}{arXiv:2005.13961}.

\bibitem{BDJ99}
Baik J., Deift P., Johansson K., On the distribution of the length of the
 longest increasing subsequence of random permutations,
 \href{https://doi.org/10.1090/S0894-0347-99-00307-0}{\textit{J.~Amer. Math.
 Soc.}} \textbf{12} (1999), 1119--1178,
 \href{http://arxiv.org/abs/math.CO/9810105}{arXiv:math.CO/9810105}.

\bibitem{B+19a}
Bailey E.C., Bettin S., Blower G., Conrey J.B., Prokhorov A., Rubinstein M.O.,
 Snaith N.C., Mixed moments of characteristic polynomials of random unitary
 matrices, \href{https://doi.org/10.1063/1.5092780}{\textit{J.~Math. Phys.}}
 \textbf{60} (2019), 083509, 26~pages,
 \href{http://arxiv.org/abs/1901.07479}{arXiv:1901.07479}.

\bibitem{B+19}
Basor E., Bleher P., Buckingham R., Grava T., Its A., Its E., Keating J.P., A
 representation of joint moments of {CUE} characteristic polynomials in terms
 of {P}ainlev\'e functions,
 \href{https://doi.org/10.1088/1361-6544/ab28c7}{\textit{Nonlinearity}}
 \textbf{32} (2019), 4033--4078,
 \href{http://arxiv.org/abs/1811.00064}{arXiv:1811.00064}.

\bibitem{BFK95}
Bergeron F., Favreau L., Krob D., Conjectures on the enumeration of tableaux of
 bounded height, \href{https://doi.org/10.1016/0012-365X(94)00148-C}{\textit{Discrete Math.}} \textbf{139} (1995), 463--468.

\bibitem{BG00}
Bergeron F., Gascon F., Counting {Y}oung tableaux of bounded height,
 \textit{J.~Integer Seq.} \textbf{3} (2000), 00.1.7, 8~pages.

\bibitem{Bo24a}
Bornemann F., Asymptotic expansions relating to the distribution of the length
 of longest increasing subsequences,
 \href{https://doi.org/10.1017/fms.2024.13}{\textit{Forum Math. Sigma}}
 \textbf{12} (2024), e36, 56~pages,
 \href{http://arxiv.org/abs/2301.02022}{arXiv:2301.02022}.

\bibitem{Bo24}
Bornemann F., A {S}tirling-type formula for the distribution of the length of
 longest increasing subsequences,
 \href{https://doi.org/10.1007/s10208-023-09604-z}{\textit{Found. Comput.
 Math.}} \textbf{24} (2024), 915--953,
 \href{http://arxiv.org/abs/2206.09411}{arXiv:2206.09411}.

\bibitem{B+20}
Bostan A., Elvey~Price A., Guttmann A.J., Maillard J.-M., Stieltjes moment
 sequences for pattern-avoiding permutations,
 \href{https://doi.org/10.37236/9402}{\textit{Electron.~J. Combin.}}
 \textbf{27} (2020), 20, 59~pages,
 \href{http://arxiv.org/abs/2001.00393}{arXiv:2001.00393}.

\bibitem{BS09}
Bostan A., Schost {\'{E}}., A simple and fast algorithm for computing
 exponentials of power series,
 \href{https://doi.org/10.1016/j.ipl.2009.03.012}{\textit{Inform. Process.
 Lett.}} \textbf{109} (2009), 754--756,
 \href{http://arxiv.org/abs/1301.5804}{arXiv:1301.5804}.

\bibitem{BW25}
Bothner T., Wei F., Asymptotic analysis of a family of {P}ainlev\'e functions
 with applications to CUE derivative moments,
 \href{http://arxiv.org/abs/2511.18118}{arXiv:2511.18118}.

\bibitem{BF25}
Byun S.-S., Forrester P.J., Progress on the study of the {G}inibre ensembles,
 \textit{KIAS Springer Ser. Math.}, Vol.~3,
 \href{https://doi.org/10.1007/978-981-97-5173-0}{Springer}, Singapore, 2025.

\bibitem{CRS06}
Conrey J.B., Rubinstein M.O., Snaith N.C., Moments of the derivative of
 characteristic polynomials with an application to the {R}iemann zeta
 function, \href{https://doi.org/10.1007/s00220-006-0090-5}{\textit{Comm.
 Math. Phys.}} \textbf{267} (2006), 611--629.

\bibitem{Da68}
Davis A.W., A system of linear differential equations for the distribution of
 {H}otelling's generalized~{$T_{0}^{2}$},
 \href{https://doi.org/10.1214/aoms/1177698313}{\textit{Ann. Math. Statist.}}
 \textbf{39} (1968), 815--832.

\bibitem{De08}
Dehaye P.-O., Joint moments of derivatives of characteristic polynomials,
 \href{https://doi.org/10.2140/ant.2008.2.31}{\textit{Algebra Number Theory}}
 \textbf{2} (2008), 31--68,
 \href{http://arxiv.org/abs/math.NT/0703440}{arXiv:math.NT/0703440}.

\bibitem{DF17}
Diaconis P., Forrester P.J., Hurwitz and the origins of random matrix theory in
 mathematics, \href{https://doi.org/10.1142/S2010326317300017}{\textit{Random
 Matrices Theory Appl.}} \textbf{6} (2017), 1730001, 26~pages,
 \href{http://arxiv.org/abs/1512.09229}{arXiv:1512.09229}.

\bibitem{Fo25}
Forrester P.J., Dualities in random matrix theory,
 \href{http://arxiv.org/abs/2501.07144}{arXiv:2501.07144}.

\bibitem{Fo93c}
Forrester P.J., Exact results and universal asymptotics in the {L}aguerre
 random matrix ensemble,
 \href{https://doi.org/10.1063/1.530883}{\textit{J.~Math. Phys.}} \textbf{35}
 (1994), 2539--2551.

\bibitem{Fo10}
Forrester P.J., Log-gases and random matrices, \textit{Lond. Math. Soc. Monogr.
 Ser.}, Vol.~34, \href{https://doi.org/10.1515/9781400835416}{Princeton
 University Press}, Princeton, NJ, 2010.

\bibitem{Fo13}
Forrester P.J., Asymptotics of spacing distributions at the hard edge for
 {$\beta$}-ensembles,
 \href{https://doi.org/10.1142/S2010326313500020}{\textit{Random Matrices
 Theory Appl.}} \textbf{2} (2013), 1350002, 21~pages.

\bibitem{Fo19}
Forrester P.J., Meet {A}ndr\'eief, {B}ordeaux 1886, and {A}ndreev, Kharkov
 1882--1883, \href{https://doi.org/10.1142/S2010326319300018}{\textit{Random
 Matrices Theory Appl.}} \textbf{8} (2019), 1930001, 9~pages,
 \href{http://arxiv.org/abs/1806.10411}{arXiv:1806.10411}.

\bibitem{Fo22a}
Forrester P.J., Joint moments of a characteristic polynomial and its derivative
 for the circular {$\beta$}-ensemble,
 \href{https://doi.org/10.2140/pmp.2022.3.145}{\textit{Probab. Math. Phys.}}
 \textbf{3} (2022), 145--170,
 \href{http://arxiv.org/abs/2012.08618}{arXiv:2012.08618}.

\bibitem{FK22}
Forrester P.J., Kumar S., Differential recurrences for the distribution of the
 trace of the {$\beta$}-{J}acobi ensemble,
 \href{https://doi.org/10.1016/j.physd.2022.133220}{\textit{Phys.~D}}
 \textbf{434} (2022), 133220, 11~pages,
 \href{http://arxiv.org/abs/2022.13322}{arXiv:2022.13322}.

\bibitem{FM23}
Forrester P.J., Mays A., Finite size corrections relating to distributions of
 the length of longest increasing subsequences,
 \href{https://doi.org/10.1016/j.aam.2022.102482}{\textit{Adv. in Appl.
 Math.}} \textbf{145} (2023), 102482, 33~pages,
 \href{http://arxiv.org/abs/2022.10248}{arXiv:2022.10248}.

\bibitem{FR09}
Forrester P.J., Rains E.M., Matrix averages relating to {G}inibre ensembles,
 \href{https://doi.org/10.1088/1751-8113/42/38/385205}{\textit{J.~Phys.~A}}
 \textbf{42} (2009), 385205, 13~pages,
 \href{http://arxiv.org/abs/0907.0287}{arXiv:0907.0287}.

\bibitem{FR12}
Forrester P.J., Rains E.M., A {F}uchsian matrix differential equation for
 {S}elberg correlation integrals,
 \href{https://doi.org/10.1007/s00220-011-1305-y}{\textit{Comm. Math. Phys.}}
 \textbf{309} (2012), 771--792,
 \href{http://arxiv.org/abs/1011.1654}{arXiv:1011.1654}.

\bibitem{FW02}
Forrester P.J., Witte N.S., Application of the {$\tau$}-function theory of
 {P}ainlev\'e equations to random matrices:~{$\rm P_V$}, {$\rm P_{III}$}, the
 {LUE}, {JUE}, and {CUE},
 \href{https://doi.org/10.1002/cpa.3021}{\textit{Comm. Pure Appl. Math.}}
 \textbf{55} (2002), 679--727,
 \href{http://arxiv.org/abs/math-ph/0201051}{arXiv:math-ph/0201051}.

\bibitem{FW06}
Forrester P.J., Witte N.S., Boundary conditions associated with the
 {P}ainlev\'e~{III{$'$}} and~{V} evaluations of some random matrix averages,
 \href{https://doi.org/10.1088/0305-4470/39/28/S13}{\textit{J.~Phys.~A}}
 \textbf{39} (2006), 8983--8995,
 \href{http://arxiv.org/abs/math.CA/0512142}{arXiv:math.CA/0512142}.

\bibitem{Fu97}
Fulton W., Young tableaux, \textit{London Math. Soc. Stud. Texts}, Vol.~35,
 Cambridge University Press, Cambridge, 1997.

\bibitem{Ge90}
Gessel I.M., Symmetric functions and {P}-recursiveness,
 \href{https://doi.org/10.1016/0097-3165(90)90060-A}{\textit{J.~Combin. Theory
 Ser.~A}} \textbf{53} (1990), 257--285.

\bibitem{Hu01}
Hughes C.P., On the characteristic polynomial of a random unitary matrix and
 the {R}iemann zeta function, Ph.D.~Thesis, {U}niversity of Bristol, 2001.

\bibitem{KS00}
Keating J.P., Snaith N.C., Random matrix theory and~{$\zeta(1/2+it)$},
 \href{https://doi.org/10.1007/s002200000261}{\textit{Comm. Math. Phys.}}
 \textbf{214} (2000), 57--89.

\bibitem{KW24a}
Keating J.P., Wei F., Joint moments of higher order derivatives of {CUE}
 characteristic polynomials~{I}: {A}symptotic formulae,
 \href{https://doi.org/10.1093/imrn/rnae063}{\textit{Int. Math. Res. Not.}}
 \textbf{2024} (2024), 9607--9632,
 \href{http://arxiv.org/abs/2307.01625}{arXiv:2307.01625}.

\bibitem{KW24b}
Keating J.P., Wei F., Joint moments of higher order derivatives of {CUE}
 characteristic polynomials~{II}: {S}tructures, recursive relations, and
 applications,
 \href{https://doi.org/10.1088/1361-6544/ad5948}{\textit{Nonlinearity}}
 \textbf{37} (2024), 085009, 54~pages,
 \href{http://arxiv.org/abs/2307.02831}{arXiv:2307.02831}.

\bibitem{K68}
Kolchin E.R., Algebraic groups and algebraic dependence,
 \href{https://doi.org/10.2307/2373294}{\textit{Amer.~J. Math.}} \textbf{90}
 (1968), 1151--1164.

\bibitem{KK09}
Kuramoto Y., Kato Y., Dynamics of one-dimensional quantum systems, Cambridge
 University Press, 2009.

\bibitem{Ma95}
Macdonald I.G., Symmetric functions and {H}all polynomials, 2nd ed., \textit{Oxford
 Math. Monogr.}, \href{https://doi.org/10.1093/oso/9780198534891.001.0001}{The
 Clarendon Press}, New York, 1995.

\bibitem{Mi75}
Migdal A.A., Recursion equations in gauge field theories, in 30 {Y}ears of the
 {L}andau {I}nstitute~-- {S}elected {P}apers, \textit{World Sci. Ser. 20st
 Century Phys.}, Vol.~11,
 \href{https://doi.org/10.1142/9789814317344_0017}{World Scientific
 Publishing}, 1996, 114--119.

\bibitem{OEIS}
{OEIS Foundation Inc.}, The on-line encyclopedia of integer sequences, 2014,
 available at \url{https://oeis.org/}.

\bibitem{RF21}
Rahman A.A., Forrester P.J., Linear differential equations for the resolvents
 of the classical matrix ensembles,
 \href{https://doi.org/10.1142/S2010326322500034}{\textit{Random Matrices
 Theory Appl.}} \textbf{10} (2021), 2250003, 43~pages,
 \href{http://arxiv.org/abs/1908.04963}{arXiv:1908.04963}.

\bibitem{Ra98}
Rains E.M., Increasing subsequences and the classical groups,
 \href{https://doi.org/10.37236/1350}{\textit{Electron.~J. Combin.}}
 \textbf{5} (1998), 12, 9~pages.

\bibitem{Ro15}
Romik D., The surprising mathematics of longest increasing subsequences,
 \textit{IMS Textb.}, Vol.~4,
 \href{https://doi.org/10.1017/CBO9781139872003}{Cambridge University Press},
 New York, 2015.

\bibitem{SZ94}
Salvy B., Zimmermann P., {GFUN}: a maple package for the manipulation of
 generating and holonomic functions in one variable,
 \href{https://doi.org/10.1145/178365.178368}{\textit{ACM Trans. Math.
 Software}} \textbf{20} (1994), 163--177.

\bibitem{S29}
Siegel C.L., {\"{U}}ber einige Anwendungen diophantischer approximationen, in
 On some applications of {D}iophantine approximations, \textit{Quad./Monogr.},
 Vol.~2, \href{https://doi.org/10.1007/978-88-7642-520-2_2}{Edizioni della
 Normale}, Pisa, 2014, 81--138.

\bibitem{SW24}
Simm N., Wei F., On moments of the derivative of {CUE} characteristic
 polynomials and the {R}iemann zeta function,
 \href{http://arxiv.org/abs/2409.03687}{arXiv:2409.03687}.

\bibitem{St80}
Stanley R.P., Differentiably finite power series,
 \href{https://doi.org/10.1016/S0195-6698(80)80051-5}{\textit{European~J.
 Combin.}} \textbf{1} (1980), 175--188.

\bibitem{St07}
Stanley R.P., Increasing and decreasing subsequences and their variants, in
 International {C}ongress of {M}athematicians. {V}ol.~{I},
 \href{https://doi.org/10.4171/022-1/21}{European Mathematical Society},
 Z\"urich, 2007, 545--579.

\bibitem{TW94b}
Tracy C.A., Widom H., Level spacing distributions and the {B}essel kernel,
 \href{https://doi.org/10.1007/BF02099779}{\textit{Comm. Math. Phys.}}
 \textbf{161} (1994), 289--309.

\bibitem{We39}
Weyl H., The classical groups: {T}heir invariants and representations,
 \href{https://doi.org/10.2307/j.ctv3hh48t}{Princeton University Press},
 Princeton, NJ, 1939.

\bibitem{Wi12}
Winn B., Derivative moments for characteristic polynomials from the {CUE},
 \href{https://doi.org/10.1007/s00220-012-1512-1}{\textit{Comm. Math. Phys.}}
 \textbf{315} (2012), 531--562,
 \href{http://arxiv.org/abs/1109.0227}{arXiv:1109.0227}.

\bibitem{Z}
Zeilberger D., A holonomic systems approach to special functions identities,
 \href{https://doi.org/10.1016/0377-0427(90)90042-X}{\textit{J.~Comput. Appl.
 Math.}} \textbf{32} (1990), 321--368.

\end{thebibliography}
\end{document}